\RequirePackage[2020-02-02]{latexrelease}

\documentclass[%
reprint,
amsmath,amssymb,
aps,
pra,
]{revtex4-2}

\makeatletter
\DeclareRobustCommand\ref{%
	\@ifstar\@refstar\T@ref
}%
\DeclareRobustCommand\pageref{%
	\@ifstar\@pagerefstar\T@pageref
}%
\makeatother

\usepackage{graphicx}
\usepackage{dcolumn}
\usepackage{bm}
\usepackage[colorlinks = true,
linkcolor = blue,
urlcolor  = blue,
citecolor = blue,
anchorcolor = black]{hyperref}  
\usepackage{physics}
\usepackage{epstopdf}
\usepackage{epsfig}
\usepackage{mathtools}

\begin{document}
	
	\preprint{APS/123-QED}
	
	\title{Atomic corrections for the unique first forbidden $\beta$ transition of $^{187}$Re}
	
	\author{O. Ni\c{t}escu$^{1,2,3}$}
	\email{ovidiu.nitescu@nipne.ro}
	\author{R. Dvornick\'y$^{1}$}
	\email{dvornicky@fmph.uniba.sk}
	
	\author{F. \v{S}imkovic$^{1,4}$}
	\email{fedor.simkovic@fmph.uniba.sk}
	
	\affiliation{$^{1}$Faculty of Mathematics, Physics and Informatics, Comenius University in Bratislava, 842 48 Bratislava, Slovakia}
	
	\affiliation{$^{2}$ International Centre for Advanced Training and Research in Physics, P.O. Box MG12, 077125 M\u{a}gurele, Romania}

	\affiliation{$^{3}$“Horia Hulubei” National Institute of Physics and Nuclear Engineering, 30 Reactorului, POB MG-6, RO-077125 Bucharest-M\u{a}gurele, Romania}


	\affiliation{ $^{4}$Institute of Experimental and Applied Physics, Czech Technical University in Prague,
		110 00 Prague, Czech Republic}
	\begin{abstract}
		
		In this paper, we reexamine one of the most promising candidates for determining the neutrino mass scale -- the unique first forbidden $\beta$ transition from $^{187}$Re($5/2^+$) to $^{187}$Os($1/2^-$). With the lowest-known ground-state to ground-state $Q$-value for a $\beta$ transition at $2.4709$ keV, rhenium's $\beta$ decay can offer insights into the neutrino mass scale puzzle. However, understanding its electron spectrum is a complex task. Besides involving a mixture of $s_{1/2}$-state and $p_{3/2}$-state electrons, the rhenium $\beta$ spectrum could be strongly influenced by various atomic corrections. In addition to our previous paper, we have incorporated finite nuclear size, diffuse nuclear surface, screening, and exchange corrections into the rhenium $\beta$ decay model. We have accounted for the last two effects within the framework of the Dirac-Hartree-Fock-Slater self-consistent method. We have discovered that both screening and exchange effects significantly alter the partial decay rates for the $s_{1/2}$- and $p_{3/2}$-state emission channels, while still maintaining the experimentally confirmed dominance of the $p_{3/2}$-state emission. The ratio between the respective decay rates has been found to be approximately $10^4$. When compared to the other corrections, the exchange effect stands out due to the modification it induces in the spectrum shape. We demonstrate that calculations with and without the exchange effect lead to entirely different shape factors for the decay spectrum. Finally, we illustrate that to preserve the linearity of the Kurie plot, it is essential to include the exchange correction in its definition. We conclude that atomic effects, especially the exchange effect, should be taken into account in current and future investigations of the neutrino mass scale from $\beta$ decays.

	\end{abstract}
	
	\maketitle
	
	
	\section{Introduction}

	An open question in particle physics is related to the absolute values of the light neutrinos masses. The information gathered from solar neutrinos, atmospheric neutrinos, nuclear power reactors, and experiments at accelerators has established that at least two of the three neutrinos have a non-zero mass. However, the result is based on neutrino oscillations, which depends on neutrino masses differences, not on their absolute values, so the absolute neutrino mass scale is uncertain. Besides the cosmological observations, which recently yielded an upper limit for the sum of the neutrino masses of $\sim 0.12$ eV \cite{CosmologyNeutrinoMassAA2022}, other puzzle pieces can be obtained from nuclear $\beta$ decay and neutrinoless double $\beta$ decay \cite{AvignoneRMP2008,EjiriPR2019,SimkovicPU2021}.

	The distortion in the endpoint measurements of the spectrum of electrons emitted in a $\beta$ decay offers a direct means of determining the values of neutrino masses, denoted as $m_k$ ($k=1,2,3$). However, the number of events emitted near the endpoint, within an interval $\Delta T_e$, is proportional to $(\Delta T_e/Q)^3$ \cite{FerriPP2015, RastislavPRC2011}. Therefore, a low $Q$-value $\beta$ transition is desirable to enhance sensitivity. Consequently, some experiments are based on the ground-state to ground-state $\beta$ decays of tritium ($^3$H) and rhenium ($^{187}$Re) with $Q$-values of $18.6$ keV \cite{MyersPRL2015} and $2.4709$ keV \cite{Qvalue187Re}, respectively. Other experiments, such as HOLMES \cite{HOLMESexperimentJLTP2016}, NuMECs \cite{NuMECsexperimentJLTP2016}, and ECHo \cite{ECHoTEPJST2017}, aim to utilize the lowest energy electron capture of $^{163}$Ho, which has a ground-state to ground-state $Q$-value of $2.833$ keV \cite{EliseevPRL2016}. Recently, there has also been a growing interest in ultra-low $Q$-value (under $1$ keV) ground-state-to-excited-state $\beta$ transitions \cite{GamageHY2019,RedshawEPJA2023,KeblbeckPRC2023}, which represent potential candidates for future neutrino mass scale determination experiments.

	The current best upper limit on effective neutrino mass, 
	\begin{equation}
		m_\beta=\sqrt{\sum_{k=1}^{3}\left|U_{ek}\right|^2m_k^2}
	\end{equation}
	was recently fixed from the tritium $\beta$ decay, measured by the KATRIN experiment \cite{KATRIN-N2022},  $m_\beta \leqslant 0.8$ eV. This limit far exceeds the previous investigations by Troitsk experiment, $m_\beta \leqslant 2.2$ eV \cite{TroitskPRD2011}, and Mainz experiment, $m_\beta \leqslant 2.3$ eV \cite{MainzEPJC2005}, also based on tritium $\beta$ decay. Even more recently, Project 8 has demonstrated that cyclotron radiation emission spectroscopy (CRES) can constrain $m_\beta \leq 155$ eV from the tritium $\beta$ spectrum using only a cm$^{3}$-scale physical detection medium \cite{Project8PRL2023}. This suggests that CRES is an attractive technique for next-generation direct neutrino mass experiments and for measuring the $\beta$ spectra in general.    
	
	While the number of events near the endpoint is favored by the $Q$-value of rhenium decay, which is nearly one order of magnitude smaller than that of tritium decay, understanding its spectrum shape is a complex task. Firstly, the emission of low-energy electrons suggests that various atomic and molecular effects could have a significant impact on the spectrum shape of the rhenium ground-state to ground-state $\beta$ transition. An example of this is the oscillations observed in the rhenium $\beta$ spectrum due to environmental fine structure from the target crystal \cite{GattiN1999}. Secondly, in contrast to the decay of tritium, which is an allowed transition, the transition $^{187}$Re($5/2^+$)$\rightarrow$$^{187}$Os($1/2^-$) is classified as a unique first forbidden transition. Consequently, its spectrum is a mixture of $s_{1/2}$-state and $p_{3/2}$-state electrons. The exceptionally low $Q$-value of rhenium decay results in a clear dominance of the $p_{3/2}$-state electron emission channel over the $s_{1/2}$-state electron emission channel. This dominance has been experimentally confirmed \cite{MARE-PRL2006}, and our previous investigation calculated the ratio between the channels to be around $10^4$ \cite{RastislavPRC2011}. 
	
	The experimental efforts to achieve sub-eV sensitivity to the neutrino mass from the rhenium $\beta$ spectrum culminated with the combination of the MANU and MIBETA groups into the Microcalorimeter Arrays for a Rhenium Experiment (MARE) \cite{NucciottiJLTP2008}. The MARE project eventually transitioned to the holmium efforts of ECHo and HOLMES \cite{FormaggioPR2021}. One possible reason for closing the project is the lack of theoretical knowledge about the rhenium $\beta$ spectrum. In contrast, the theoretical description of tritium $\beta$ decay involves many more corrections, including screening, exchange with atomic electrons, finite nuclear size effects, radiative corrections, recoil effects, etc. \cite{WilkinsonNPA1991, Kleesiek2019}.

	In this paper, we investigate the $\beta$ decay of $^{187}$Re, incorporating all relevant corrections to its spectrum. We begin with the same relativistic wave functions for the emitted electrons as used in our previous study \cite{RastislavPRC2011}. Next, we enhance the precision of the theoretical rhenium decay spectrum by including finite nuclear size, diffuse nuclear surface, and screening corrections. The latter is calculated using the self-consistent Dirac-Hartree-Fock-Slater description of the atomic bound electrons surrounding the electron emitted during rhenium decay. We have observed significant differences in the decay rates for both emission channels compared to our previous paper, but negligible modifications to the spectral shape resulting from the aforementioned corrections.

	The primary focus of this paper lies in the inclusion of the so-called exchange correction, which considers the possible interchange between emitted electrons and the atomic bound electrons. It turns out that the exchange correction not only alters the decay rates for the $p_{3/2}$- and $s_{1/2}$-state channels but also affects the shapes of the spectra for these channels. While the most significant shape modification occurs in the low-energy region of the total spectrum, the exchange correction remains substantial near the endpoint, potentially impacting the analysis for neutrino mass scale determination from rhenium $\beta$ decay. When studying the deviation of the rhenium spectrum from an allowed one, we discover that the exchange correction strongly transforms the shape factor, shifting it from an increasing linear behavior to a decreasing quadratic one. We present the best-fit parameters in both cases. To preserve the linearity of the Kurie plot in the scenario of zero effective neutrino mass, it is necessary to revise its definition to incorporate the exchange correction. We illustrate how varying effective neutrino masses impact the Kurie plots near the endpoint of the $\beta$ decay of $^{187}$Re. While this paper is primarily aimed on the rhenium $\beta$ decay, we emphasize that the corrections introduced here could potentially have a significant effect on any low and ultra-low $Q$-value $\beta$ transition.

	\section{Electron Wave functions}
	\subsection{Continuum states}
	\label{sec:ApproximationSchemes}
	
	The relativistic wave function for an electron emitted from a $\beta$ decay in a continuum state can be expanded in spherical waves,
	
	\begin{eqnarray}
		\psi(E_e,\boldsymbol{r})=&&\psi^{(s_{1/2})}(E_e,\boldsymbol{r})+\psi^{(p_{1/2})}(E_e,\boldsymbol{r})\nonumber\\&&+\psi^{(p_{3/2})}(E_e,\boldsymbol{r})+\dots,
	\end{eqnarray} 
	where $E_e$ is the total electron energy. The superscript in the spherical waves represents the orbital and total angular momentum $(\ell_j=0_{1/2},1_{1/2},1_{3/2},...)$ written in the spectroscopic notation $(\ell_j=s_{1/2},p_{1/2},p_{3/2},...)$. The relevant spherical waves for the first-unique forbidden $\beta$ transition of $^{187}$Re are as follows \cite{RastislavPRC2011}:
	
	\begin{widetext}
		\begin{eqnarray}
			\psi^{(s_{1/2})}(E_e,\boldsymbol{r})=\begin{pmatrix}
				g_{-1}(E_e,r)\chi_\mu\\
				f_{+1}(E_e,r)(\boldsymbol{\sigma}\cdot\hat{\boldsymbol{p}})\chi_\mu\end{pmatrix},\hspace{0.4cm}
			\psi^{(p_{3/2})}(E_e,\boldsymbol{r})=i\begin{pmatrix}
				g_{-2}(E_e,r)\left[3(\hat{\boldsymbol{r}}\cdot\hat{\boldsymbol{p}})-(\boldsymbol{\sigma}\cdot\hat{\boldsymbol{r}})(\boldsymbol{\sigma}\cdot\hat{\boldsymbol{p}})\right]\chi_\mu\\
				f_{+2}(E_e,r)\left[(\hat{\boldsymbol{r}}\cdot\hat{\boldsymbol{p}})(\boldsymbol{\sigma}\cdot\hat{\boldsymbol{p}})-(\boldsymbol{\sigma}\cdot\hat{\boldsymbol{r}})\right]\chi_\mu
			\end{pmatrix},
			\label{eq:wideeq}
		\end{eqnarray}
	\end{widetext}
	where $\hat{\boldsymbol{p}}=\boldsymbol{p_e}/p_e$ is defined by the momentum of the electron, $p_e=\sqrt{E_e^2-m_e^2}$. Here, $\boldsymbol{r}$ stands for the position vector of the electron, $r=\left|\boldsymbol{r}\right|$ and $\hat{\boldsymbol{r}}=\boldsymbol{r}/r$. Throughout the paper, we are using the units $\hbar=c=1$.

	The functions $g_\kappa(E_e,r)$ and $f_\kappa(E_e,r)$ are the large- and small-component radial wave functions, respectively, which satisfy the following system of coupled differential equations \cite{RoseBook1961}
	\begin{eqnarray}
		\label{eq:radialEquations}
		\left(\frac{d}{dr}+\frac{\kappa+1}{r}\right)g_\kappa-(E_e-V(r)+m_e)f_\kappa=0,\nonumber\\
		\left(\frac{d}{dr}-\frac{\kappa-1}{r}\right)f_\kappa+(E_e-V(r)-m_e)g_\kappa=0,
	\end{eqnarray}
	where $V(r)$ is the atomic potential of the final system.
	
	The relativistic quantum number, $\kappa$, takes positive and negative integer values, and specifies both the total angular momentum, $j$, and the orbital angular momentum, $\ell$, by
	\begin{eqnarray}
		j=\left|\kappa\right|-1/2, \hspace{0.2cm} \ell=\begin{cases}
			\kappa  &\quad \text{if } \kappa>0,\\
			\left|\kappa\right|-1  &\quad \text{if } \kappa<0.\\
		\end{cases}
	\end{eqnarray}
	
	The large- and small-component radial functions satisfy, for large values of $p_er$, the following asymptotic conditions
	\begin{eqnarray}
		&&\begin{Bmatrix}
			g_{\kappa}(E_e,r)\\
			f_{\kappa}(E_e,r)
		\end{Bmatrix}\\
		&&\sim \frac{1}{p_er}
		\begin{Bmatrix}
			\sqrt{\frac{E_e+m_e}{2E_e}} \sin (p_er-l\frac{\pi}{2}+\delta_{\kappa}-\eta \ln (2p_er))\\
			\sqrt{\frac{E_e-m_e}{2E_e}} \cos (p_er-l\frac{\pi}{2}+\delta_{\kappa}-\eta \ln (2p_er))
		\end{Bmatrix}\nonumber
	\end{eqnarray}
	where $\delta_{\kappa}$ is the phase shift and $\eta=\alpha Z E_e/p_e$ is the Sommerfeld parameter. Here, $Z$ is the atomic number of the final nucleus and $\alpha$ is the fine-structure constant. The energy-dependent normalization ensures the correct behavior of the Fermi functions (see Sec.~\ref{sec:RheniumDecay}) when the electrostatic interaction is switched off, i.e., $V(r)=0$. 
	
	To obtain the electron continuum states in the potential of the final positive ion, $^{187}$Os$^+$, we employ the RADIAL subroutine package \cite{SalvatCPC2019}. A comprehensive manual of the package can be found in the Supplementary Material of \cite{SalvatCPC2019}. In what follows, we briefly present different approximation schemes for the calculation of $g_{\kappa}(E_e,r)$ and $f_{\kappa}(E_e,r)$.
	
	\textit{The approximation scheme A.} We assume the final nucleus as a uniformly charged sphere, generating the following potential,
	\begin{equation}
		\label{eq:UniformChargedSpherePotential}
		V(r)=\begin{cases}
			-\frac{\alpha Z}{r}  &\quad \text{for } r\ge R,\\
			-\frac{\alpha Z}{2R}\left[3-\left(\frac{r}{R}\right)^2\right]  &\quad \text{for } r<R.\\
		\end{cases}
	\end{equation}
	Here, $R$ is the radius of the final nucleus, $R=r_0A^{1/3}$ with $r_0=1.2$ fm.  By keeping the lowest power of the expansion of $r$, the radial wave functions for the $s_{1/2}$ wave and $p_{3/2}$ wave states are given by \cite{DoiPTPS1985}
	
	\begin{equation}
		\begin{pmatrix}
			g_{-1}(E_e,r)\\
			f_{+1}(E_e,r)
		\end{pmatrix}=
		\begin{pmatrix}
			A_{-1}\\
			A_{+1}
		\end{pmatrix},
	\end{equation}
	and 
	
	\begin{equation}
		\begin{pmatrix}
			g_{-2}(E_e,r)\\
			f_{+2}(E_e,r)
		\end{pmatrix}=
		\frac{p_er}{3}\begin{pmatrix}
			A_{-2}\\
			A_{+2}
		\end{pmatrix},
	\end{equation}
	respectively.
	
	The normalization constant can be expressed in a good approximation as
	
	\begin{equation}
		A_{\pm k}\simeq\sqrt{F_{k-1}(Z,E_e)}\sqrt{\frac{E_e\mp m_e}{2E_e}}
	\end{equation} 
	where $k=\left|\kappa\right|$ and the Fermi function $F_{k-1}(Z,E_e)$ is given by
	
	\begin{eqnarray}
		F_{k-1}(Z,E_e)&=&\left[\frac{\Gamma(2k+1)}{\Gamma(k)\Gamma(2\gamma_k+1)}\right]^2(2pR)^{2(\gamma_k-k)}e^{\pi\eta} \nonumber\\ 
		&&\times \left|\Gamma(\gamma_k+i\eta)\right|^2.
	\end{eqnarray}
	The remaining quantity is given by
	\begin{align}
		\begin{aligned}
			\gamma_k&=\sqrt{k^2-(\alpha Z)^2},
		\end{aligned}
	\end{align} 
	and $\Gamma(z)$ is the Gamma function. We mention that this approximation scheme, was also used in our previous investigation \cite{RastislavPRC2011}.
	
	\textit{The approximation scheme B.} In this approximation scheme, we consider the case, where the final nucleus generates a point-like potential, $V(r)=-\alpha Z /r$. The radial wave functions can be expressed analytically as \cite{BeresteckijBook1989}
	\begin{align}
		\begin{aligned}
			\label{eq:WFPointAnalytical}
			g_{\kappa}(E_e,r)=\frac{\kappa}{k}\frac{1}{pr}\sqrt{\frac{E_e+m_e}{2E_e}}\frac{\left|\Gamma(1+\gamma_k+i \eta)\right|}{\Gamma(1+2\gamma_k)}(2pr)^{\gamma_k}e^{\pi\eta/2}\\
			\times \Im \{e^{i(pr+\zeta)} \hspace{0.01cm}_1F_1(\gamma_k-i\eta,1+2\gamma_k,-2ipr)\},\\
			f_{\kappa}(E_e,r)=\frac{\kappa}{k}\frac{1}{pr}\sqrt{\frac{E_e-m_e}{2E_e}}\frac{\left|\Gamma(1+\gamma_k+i \eta)\right|}{\Gamma(1+2\gamma_k)}(2pr)^{\gamma_k}e^{\pi\eta/2}\\
			\times \Re \{e^{i(pr+\zeta)} \hspace{0.01cm}_1F_1(\gamma_k-i\eta,1+2\gamma_k,-2ipr)\},\\
		\end{aligned}
	\end{align} 
	with
	\begin{equation}
		e^{i\zeta}=\sqrt{\frac{\kappa-i\eta m_e/E_e}{\gamma_k-i\eta}}.
	\end{equation}
	Here, $_1 F_1(a,b,z)$ is the confluent hypergeometric function. We mention that the numerical solutions from the RADIAL package, with the input $rV(r)=-\alpha Z$, are equivalent to the analytical solutions presented in Eq.~(\ref{eq:WFPointAnalytical}), if we fix $r$ on the nuclear surface.

	\textit{The approximation scheme C.} We consider the final nucleus as a sphere filled with protons following a Fermi distribution  \cite{HahnPR1956}    
	\begin{eqnarray}
		\rho_p(r)=\frac{\rho_0}{1+e^{\left(r-c_{\text{rms}}\right)/a}},
	\end{eqnarray}
	where we chose for the half-way radius $c_{\text{rms}}=1.07A^{1/3}=6.118$ fm and for the surface thickness $a=0.546$ fm. $\rho_0$ is determined from the normalization to $Z$.
	
	Thus, the electrostatic interaction of an electron at $r$ with the final nucleus is described by  
	\begin{eqnarray}
		\label{eq:nuclearPotential}
		V_{\text{nuc}}(r)=-\alpha\int\frac{\rho_p(r')}{|\boldsymbol{r}-\boldsymbol{r'}|}d\boldsymbol{r'}.
	\end{eqnarray}
	
	\textit{The approximation scheme D.} In our final approximation scheme, we consider for the final positive ion, $^{187}$Os$^+$, the following electrostatic potential,
	\begin{eqnarray}
		\label{eq:PotentialSchemeD}
		V(r)=V_{\text{nuc}}(r)+V_{\text{el}}(r)+V^{\text{Slater}}_{\text{ex}}(r),
	\end{eqnarray}
	which is a sum of the nuclear, electronic and exchange potentials. The nuclear potential, $V_{\text{nuc}}(r)$, is presented in the approximation scheme C, i.e. Eq.~(\ref{eq:nuclearPotential}). The electronic potential, $V_{\text{el}}(r)$ , describes the interaction energy of an electron at $r$ with the atomic cloud, and it is found from integrating over the volume of the atomic electron density, $\rho(r)$, 
	\begin{eqnarray}
		V_\text{el}(r)=\alpha\int\frac{\rho(r')}{|\boldsymbol{r}-\boldsymbol{r'}|}d\boldsymbol{r'}.
	\end{eqnarray} 
	Considering Slater's approximation \cite{SlaterPR1951}, we can write the exchange potential in terms of the atomic electron density,
	\begin{eqnarray}
		\label{eq:SlaterExchange}
		V_{\text{ex}}^{\text{Slater}}(r)=-\frac{3}{2}\alpha	\left(\frac{3}{\pi}\right)^{1/3}\left[\rho(r)\right]^{1/3}.
	\end{eqnarray}
	The atomic electronic density, $\rho(r)$, is discussed in the following section, where we present the calculation for the electron bound states.
	
	Our choice to include a local exchange component in the potential for the continuum states can be arguable, but, as motivated later, it ensures the orthogonality between the continuum states of the emitted electron and the bound states of the atomic electrons of $^{187}$Os$^+$. Orthogonal continuum and bound electron wave functions for the final nucleus are crucial ingredients for correctly calculating the exchange correction \cite{HarstonPRA1992,PyperPRSLA1998,NitescuPRC2023}.

	\begin{figure}[h]
		\includegraphics[width=0.48\textwidth]{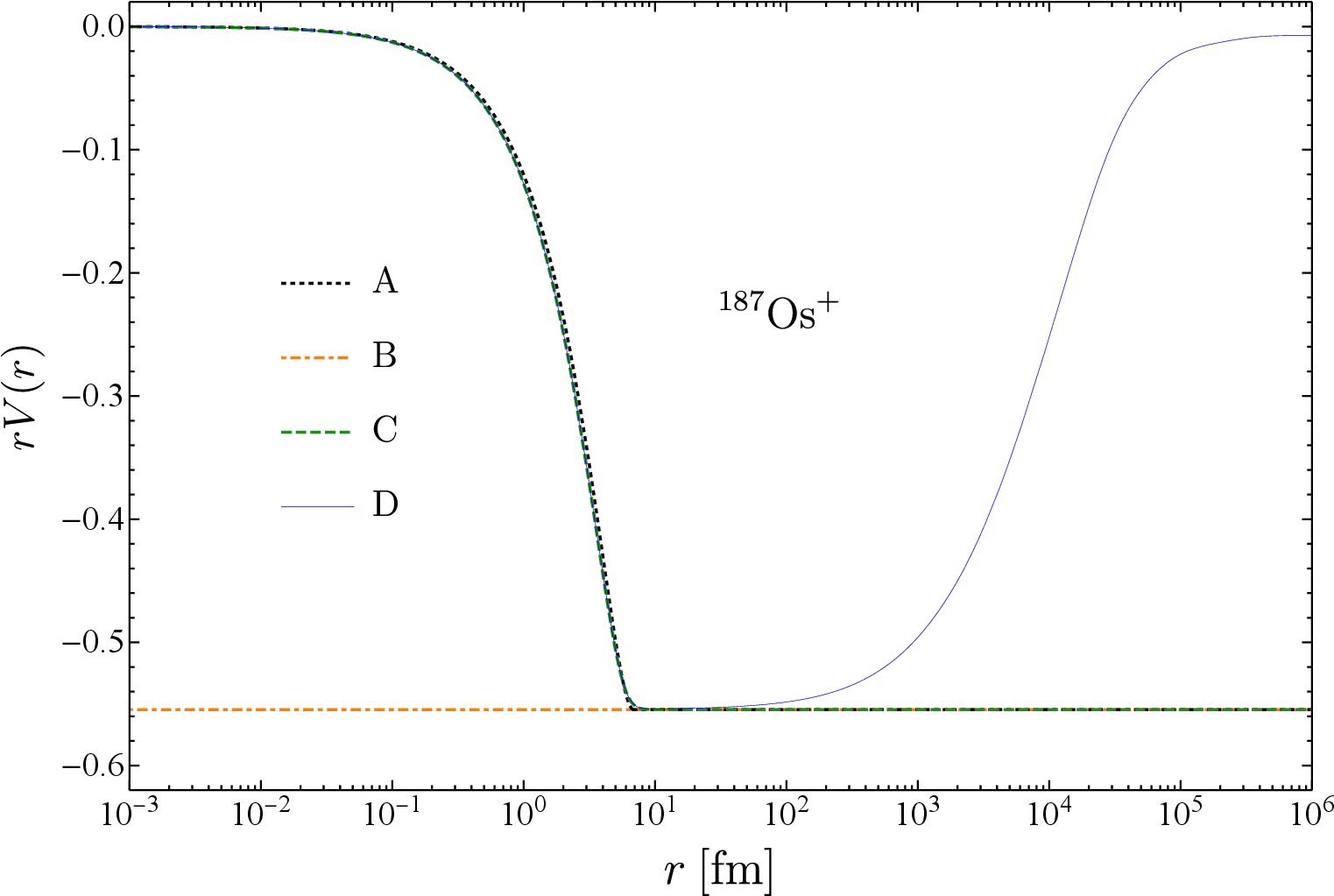}
		\caption{The electrostatic potential for $^{187}$Os$^+$ as function of $r$, where the emitted electron is located, in four different approximation schemes: (A) The final nucleus as an uniform charged sphere. (B) A point-like final nucleus. (C) The final nucleus as a charged sphere filled with protons following a Fermi distribution. (D) The same as the preceding case but the atomic electron screening is taken into account.      \label{fig:PotentialsFor187OsPositiveIon}} 
	\end{figure}

	In Fig.~\ref{fig:PotentialsFor187OsPositiveIon}, we plot all quantities $rV(r)$ as functions of $r$, for each approximation scheme considered above. In the simplest case of a point-like nucleus (scheme B), we can see that $rV(r)=-76\alpha$ is a straight line. In the most complex case (scheme D), where we include the finite nuclear size, diffuse nuclear surface, and the atomic screening corrections, for large values of $r$, $V(r)=-\alpha/r$. It is an expected asymptotic result because the potential describes the interaction of the emitted electron with a positive ion of charge $+1$. We can see that when moving from a uniformly charged sphere (scheme A) to one filled with protons following a Fermi distribution (scheme C), the difference between potentials is slightly visible. Still, the electron wave functions A are approximated by keeping the lowest power in the expansion in $r$. By solving the radial Dirac equation numerically for the potentials A and C, we confirm that there are no differences between the wave functions due to the diffuse nuclear surface correction. Thus, any differences between the electron wave functions A and C are associated with missing terms in the expansion.

	\begin{figure*}
		\includegraphics[width=0.7\textwidth]{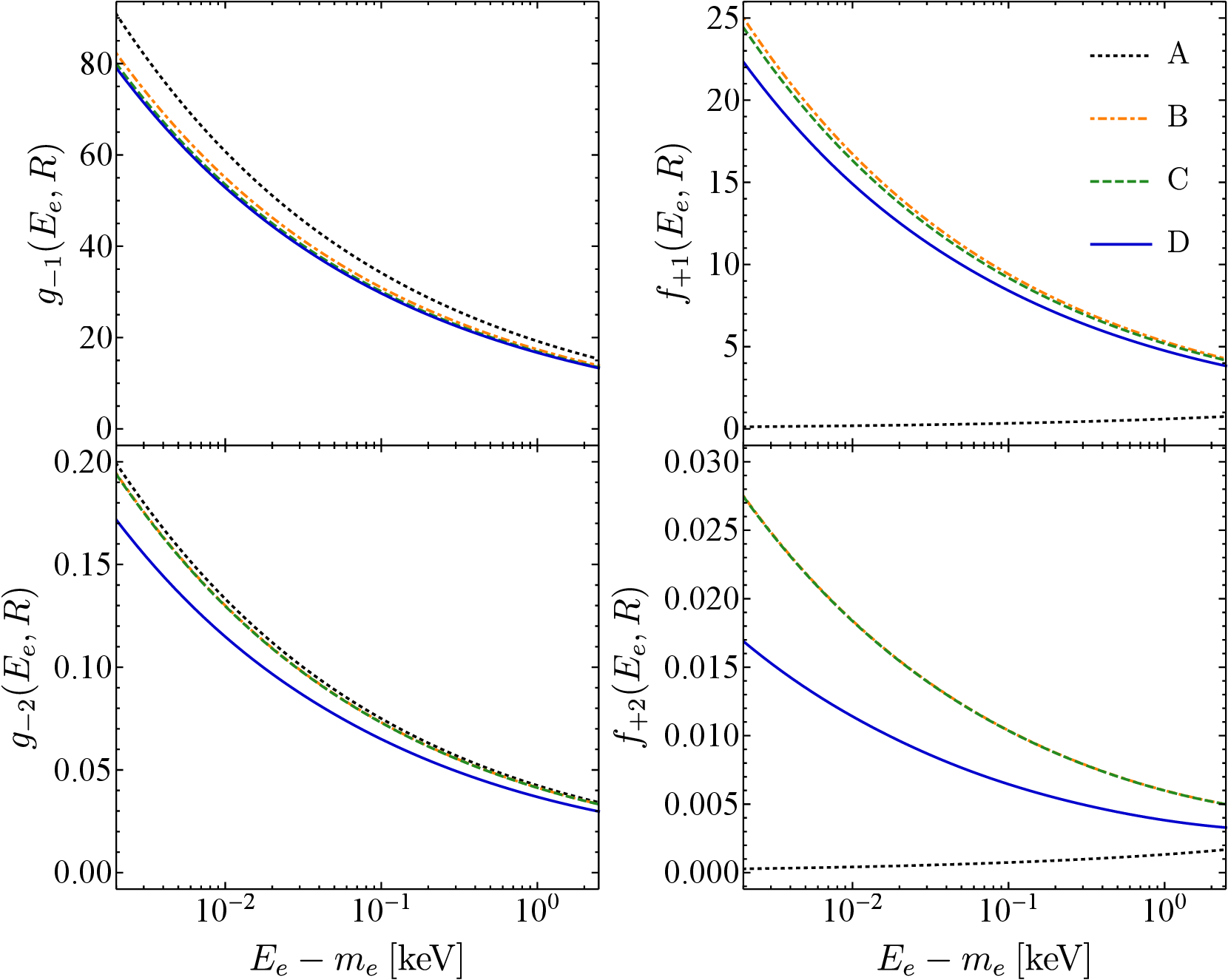}
		\caption{The radial wave functions for an electron emitted from the $\beta$ decay of $^{187}$Re, evaluated on the nuclear surface of the final nucleus $ R $, as functions of the electron kinetic energy $E_e-m_e$. In the top panel, the components of on electron emitted in $s_{1/2}$ state, $g_{-1}(E_e,r)$ and $f_{+1}(E_e,r)$, and in the bottom panel, the components associated with the emission in $p_{3/2}$ state, $g_{-2}(E_e,r)$ and $f_{+2}(E_e,r)$. We consider four different approximation schemes in the evaluation of the radial wave functions (see text for details).   \label{fig:WaveFunctions}} 
	\end{figure*}

	In Fig.~\ref{fig:WaveFunctions}, we depict the radial wave functions of an electron in the $s_{1/2}$ wave state, $g_{-1}(E_e,r)$ and $f_{+1}(E_e,r)$ (top panel), and in the $p_{3/2}$ wave state, $g_{-2}(E_e,r)$ and $f_{+2}(E_e,r)$ (bottom panel), evaluated on the nuclear surface of the final nucleus, $r=R$, and as functions of the electron kinetic energy, $E_e-m_e$. We present the results for the $\beta$ emitter $^{187}$Re using all the approximation schemes. In the case of the large component radial wave functions, $g_{-1}(E_e,r)$ and $g_{-2}(E_e,r)$, there are no shape deviations between different approximation schemes. Their amplitudes progressively decrease from scheme A to D. The missing terms in the expansion in $r$ from scheme A strongly influence the small component radial wave functions, $f_{+1}(E_e,r)$ and $f_{+2}(E_e,r)$. Instead of overlying the wave functions C, they are almost zero and constants for any electron kinetic energy. As a final remark, the effect of the atomic screening for an electron emitted in the $p_{3/2}$ wave state is substantial, as can be seen in the wave functions D. The emission of electrons in the $p_{3/2}$ wave state is dominant for the $\beta$ decay of $^{187}$Re, so we expect a considerable deviation when the screening correction is switched on.

	\subsection{Bound states}
	For atomic bound states ($E_e<m_e$), each discrete energy level is characterized by its quantum number $\kappa$, its principle quantum number $n$, its binding energy $\epsilon_e^{n\kappa}$, and its total energy $E_e^{n\kappa}=m_e-\left|\epsilon_e^{n\kappa}\right|$. The relativistic wave function for a bound electron in a spherically symmetric atomic potential, $V(r)$, can be written as \cite{RoseBook1961}
	\begin{eqnarray}
		\psi_{n\kappa m}(\boldsymbol{r})=\begin{pmatrix}
			g_{n,\kappa}(r)\Omega_{\kappa, m}(\hat{\boldsymbol{r}})\\
			i f_{n,\kappa}(r)\Omega_{-\kappa, m}(\hat{\boldsymbol{r}})
		\end{pmatrix},
	\end{eqnarray} 
	Here, the spherical spinors, $\Omega_{\kappa, m}(\hat{\boldsymbol{r}})$, are defined by \cite{RoseBook1995,VarshalovichBook1988}
	\begin{eqnarray}
		\Omega_{\kappa, m}(\hat{\boldsymbol{r}})=\sum_{\mu=\pm1/2}\braket{\ell,\frac{1}{2},m-\mu,\mu}{j,m}Y_{\ell,m-\mu}(\hat{\boldsymbol{r}})\chi_\mu
	\end{eqnarray} 
	where $\braket{j_1,j_2,m_1,m_2}{j,m}$ are Clebsch-Gordan coefficient, $Y_{\ell,m}(\hat{\boldsymbol{r}})$ are the spherical harmonics, and $\chi_\mu$ are the usual Pauli spinors.
	
	The large- and small-component radial functions for bound orbitals are referred as $g_{n,\kappa}(r)$ and $f_{n,\kappa}(r)$, respectively, and are obeying,  
	\begin{eqnarray}
		\left(\frac{d}{dr}+\frac{\kappa+1}{r}\right)g_{n,\kappa}-(E^{n\kappa}_e-V(r)+m_e)f_{n,\kappa}=0,\nonumber\\
		\left(\frac{d}{dr}-\frac{\kappa-1}{r}\right)f_{n,\kappa}+(E^{n\kappa}_e-V(r)-m_e)g_{n,\kappa}=0,
	\end{eqnarray}
	
	For the calculation of the bound orbitals, we employed the program DHFS.f, included in the RADIAL subroutine package \cite{SalvatCPC2019}. It solves the Dirac-Hartree-Fock-Slater (DHFS) equations for the ground-state electron configuration of neutral atoms or positive ions with $N_e$ bound electrons and $Z$ protons in the nucleus. The procedure uses almost the same potential as described in Eq.~(\ref{eq:PotentialSchemeD}), but with the difference that the Latter's tail correction \cite{LatterPR1955} is imposed for the exchange potential. The electron density, $\rho(r)$, entering the electronic and exchange potentials, is obtained self-consistently \cite{LibermanCPC1971,LibermanPR1965}. The procedure starts with an approximate electron density obtained from the Moli\`ere parametrization of the Thomas-Fermi potential \cite{MoliereZNA1947}. Then, the electron density is renewed iteratively from the obtained bound orbitals until neither the DHFS potential nor the binding energies change. We call the procedure from the program DHFS.f a true DFHS self-consistent method.

	\begin{table}[h]
		\caption{\label{tab:BindingEnergies}%
			Binding energies for neutral atom $^{187}$Re in eV. In the first column, we indicate all occupied shells using the spectroscopic notation \cite{SalvatCPC2019,FroeseFischerCPC2018}. In the second and third columns, we present the true DHFS self-consistent method binding energies and the results obtained with the modified DHFS self-consistent method, respectively. In the last column, we present the experimental values taken from \cite{CarlsonBook1975}.  
		}
		\begin{ruledtabular}
			\begin{tabular}{cccc}
				\textrm{Orbital} $(n\ell_{j})$&
				\text{$\epsilon_e^{n\kappa}$ (true)} &
				\text{$\epsilon_e^{n\kappa}$} (modified)&
				\text{$\epsilon_e^{n\kappa}\textrm{(exp)}$ \cite{CarlsonBook1975}}\\
				\colrule
				$1s_{1/2}$ & -71857.5 &-71857.5&  $ -71681\pm2 $\\
				$2s_{1/2}$ & -12508.4 &-12508.4&  $ -12532\pm2 $\\
				$2p_{1/2}$ & -11993.7 & -11993.7& $ -11963\pm2 $\\
				$2p_{3/2}$ & -10537.7 & -10537.7& $ -10540\pm2 $\\
				$3s_{1/2}$ & -2911.9 & -2911.9& $ -2937\pm2 $\\
				$3p_{1/2}$ & -2677.7 & -2677.7& $ -2686\pm2 $\\
				$3p_{3/2}$ & -2360.0 & -2360.0& $ -2371\pm2 $\\
				$3d_{3/2}$ & -1961.4 & -1961.4& $ -1953\pm2 $\\
				$3d_{5/2}$ & -1891.9 & -1891.9& $ -1887\pm2 $\\
				$4s_{1/2}$ & -615.7 & -615.7& $ -629\pm2 $\\
				$4p_{1/2}$ & -516.7 & -516.7& $ -522\pm2 $\\
				$4p_{3/2}$ & -442.2 & -442.2& $ -450\pm2 $\\
				$4d_{3/2}$ & -277.8 & -277.8& $ -278\pm2 $\\
				$4d_{5/2}$ & -263.9 & -263.9& $ -264\pm2 $\\
				$5s_{1/2}$ & -91.2 & -91.2& $ -86\pm2 $\\
				$5p_{1/2}$ & -60.6 & -60.6 & $ -56\pm2 $\\
				$4f_{5/2}$ & -55.0 & -55.0& $ -47\pm2 $\\
				$4f_{7/2}$ & -52.3 & -52.3& $ -45\pm2 $\\
				$5p_{3/2}$ & -49.0 & -49.0& $ -45\pm2 $\\
				$5d_{3/2}$ & -9.28 & -9.24& $ -9.6\pm1 $\\
				$5d_{5/2}$ & -8.24 & -8.20& $ -9.6\pm1 $\\
				$6s_{1/2}$ & -7.98 & -7.67& $ -7.9\pm1 $\\
				
			\end{tabular}
		\end{ruledtabular}
	\end{table}   
	
	Because we want to ensure orthogonality between continuum and bound states, we take the last iterated atomic electron density, $\rho(r)$, obtained from the program DHFS.f, and construct the potential from the approximation scheme D, i.e., Eq.~(\ref{eq:PotentialSchemeD}). When solving for bound states in this potential, we call the procedure a modified DHFS self-consistent method. More about the differences between the true and modified DHFS self-consistent frameworks can be found in our previous paper about the exchange correction for allowed $\beta$ decay \cite{NitescuPRC2023}.

	In Table~\ref{tab:BindingEnergies}, we compare the binding energies obtained with the true and modified DHFS methods with the experimental values for the neutral atom $^{187}$Re. We can see that the deviation from the conventional (true) DHFS method subtly influences the binding energies of the last three occupied orbitals. 
	
	\section{First unique forbidden \\$\beta$ decay of $^{187}\text{Re}$}
	\label{sec:RheniumDecay}
	
	The $\beta$ transition from the ground-state $5/2^+$ of $^{187}$Re to the ground-state $1/2^-$ of $^{187}$Os is classified as a first unique forbidden $\beta$ transition. Due to the difference in the angular momentum and parity, $\Delta J^\pi=2^-$, the electron and neutrino are emitted, respectively, in $p_{3/2}$ and $s_{1/2}$ states or vice versa. An interesting feature of  first unique forbidden $\beta$ decay of $^{187}$Re is that due to its low transition energy, $Q=2.4709$ keV \cite{Qvalue187Re}, the emission of $p_{3/2}$ state electrons is favored with four orders of magnitude then the emission of $s_{1/2}$ state electrons \cite{RastislavPRC2011}. The theoretical differential decay rate is a sum of two contributions associated
	with emission of the $s_{1/2}$ and the $p_{3/2}$ state electrons, 
	\begin{eqnarray}
		\frac{d {\Gamma}}{dE_e} &&= \frac{d {\Gamma}^{p_{3/2}}}{dE_e} 
		+ \frac{d {\Gamma}^{s_{1/2}}}{dE_e} \nonumber\\
		&&= \sum_{k=1}^3 |U_{ek}|^2~\frac{G_F^2 V_{ud}^2}{2 \pi^3}~ p_e E_e  (E_0 - E_e)
		\nonumber\\
		&& \times \sqrt{(E_0 - E_e)^2-m^2_{k}}  ~\theta(E_0-E_e-m_k)\nonumber\\ 
		&& \times \left[ B^{{p_{3/2}}}(E_e,p_\nu) + B^{s_{1/2}}(E_e,p_\nu)~\right]
		\label{eq.sp1}
	\end{eqnarray}
	where
	\begin{eqnarray}
		&& B^{s_{1/2}} = \frac{1}{2}~g^2_A\times\\ 
		&&~ \left(|\langle {\rm f} \Vert \sum_n
		\tau^+_n g_{-1}(E_e,r)~j_1(p_\nu r)~
		\{ \bm{\sigma}_{n}^{~} \otimes \bm{\hat{r}}_{n}^{~} \}_{2}
		\Vert{\rm i}\rangle|^2.\label{nme}\right.\nonumber\\
		&&+~ \left.|\langle {\rm f} \Vert \sum_n
		\tau^+_n f_{+1}(E_e,r)~j_1(p_\nu r)~
		\{ \bm{\sigma}_{n}^{~} \otimes \bm{\hat{r}}_{n}^{~} \}_{2}
		\Vert {\rm i}\rangle|^2.\label{nme}\right),\nonumber
	\end{eqnarray}
	and,
	\begin{eqnarray}
		&& B^{p_{3/2}} = \frac{1}{2}~g^2_A\times\\ 
		&&~ \left(|\langle {\rm f} \Vert \sum_n
		\tau^+_n g_{-2}(E_e,r)~j_0(p_\nu r)~
		\{ \bm{\sigma}_{n}^{~} \otimes \bm{\hat{r}}_{n}^{~} \}_{2}
		\Vert{\rm i}\rangle|^2.\label{nme}\right.\nonumber\\
		&&+~ \left.|\langle {\rm f} \Vert \sum_n
		\tau^+_n f_{+2}(E_e,r)~j_0(p_\nu r)~
		\{ \bm{\sigma}_{n}^{~} \otimes \bm{\hat{r}}_{n}^{~} \}_{2}
		\Vert {\rm i}\rangle|^2.\label{nme}\right).\nonumber
	\end{eqnarray}
	Here,  $G_F$ is the Fermi constant and $V_{ud}$ is the element of the
	Cabbibo-Kobayashi-Maskawa (CKM) matrix. $E_0$
	is maximal endpoint energy (in
	the case of zero neutrino mass) of the electron.
	$p_\nu = \sqrt{(E_0 - E_e)^2-m^2_{k}}$ is the neutrino momentum and $\theta(x)$ is a theta (step) function. $g_A$ denotes an axial-vector 
	coupling constant. $|i\rangle$ ($|f\rangle$) is the initial (final)
	state of $^{187}$Re ($^{187}$Os) with $J^\pi = 5/2^+$ ($J^\pi = 1/2^-$). 
	$\mathbf{r}_{n}^{~}$ ($r = |\mathbf{r}|$ and $\bm{\hat{r}} = \mathbf{r}/r$) is a coordinate of
	the ${\rm n}$-th nucleon.

	The differential decay rate in Eq.~(\ref{eq.sp1}) depends on four different
	squared matrix elements incorporated in the $B^{s_{1/2}}$ and $B^{p_{1/2}}$ terms. For the purpose of factorization of calculation of squared nuclear matrix element and phase-space factor, the large- and small-component electron radial functions are approximated as follows \cite{RastislavPRC2011}:
	\begin{eqnarray}
		{g}_{-1}(E_e,r) &\simeq& \frac{{g}_{-1}(E_e,R)}{j_0(p_eR)}~ j_0(p_er)\simeq {g}_{-1}(E_e,R),\nonumber\\
		{f}_{+1}(E_e,r) &\simeq& \frac{{f}_{+1}(E_e,R)}{j_0(p_eR)}~ j_0(p_er)\simeq {f}_{+1}(E_e,R),\nonumber\\
		{g}_{-2}(E_e,r) &\simeq& \frac{{g}_{-2}(E_e,R)}{j_1(p_eR)}~ j_1(p_er)\simeq\frac{r}{R}{g}_{-2}(E_e,R),\nonumber\\
		{f}_{+2}(E_e,r) &\simeq& \frac{{f}_{+2}(E_e,R)}{j_1(p_eR)}~ j_1(p_er)\simeq\frac{r}{R}{f}_{+2}(E_e,R),
	\end{eqnarray}
	where $R$ is a nuclear radius. For bound states, required in the atomic exchange correction (see Sec.~\ref{sec:ExchnageCorrection}), the same approximation holds leading to $g_{n,-1}(r)\simeq g_{n,-1}(R)$, $f_{n,1}(r)\simeq f_{n,1}(R)$, $g_{n,-2}(r)\simeq (r/R)g_{n,-2}(R)$ and $f_{n,2}(r)\simeq (r/R)f_{n,2}(R)$. 
	For continuum states, only the leading terms in the expansion of spherical Bessel functions
	$j_0(p_er)$ and $j_1(p_er)$  were considered. In this way one ends up with the energy
	distribution, which depends only on a single squared matrix element \cite{RastislavPRC2011}:
	\begin{eqnarray}
		\frac{d {\Gamma}}{dE_e} &=& \frac{d {\Gamma}^{p_{3/2}}}{dE_e} 
		+ \frac{d {\Gamma}^{s_{1/2}}}{dE_e} \nonumber\\
		&=& \sum_{k=1}^3 |U_{ek}|^2~ 
		\frac{G_F^2 V_{ud}^2}{2 \pi^3}~
		B ~R^2 ~p_e~ E_e ~(E_0 - E_e)
		\nonumber\\ 
		&\times&  \frac{1}{3}
		\left[ F_{1}(Z,E_e)p_e^2 + F_{0}(Z,E_e)((E_0 - E_e)^2-m^2_{k})
		\right]
		\nonumber\\
		&& ~~\times \sqrt{(E_0 - E_e)^2-m^2_{k}}~~ 
		\theta(E_0-E_e-m_k), 
		\label{eq.sp}
	\end{eqnarray}
	with
	\begin{eqnarray}
		B &=& \frac{g_A^2}{6 R^2} \label{nme}
		|\langle {\rm f} \Vert \sum_n
		\tau^+_n 
		\{ \bm{\sigma}_{n}^{~} \otimes \bm{r}_{n}^{~} \}_{2}
		\Vert {\rm i}\rangle|^2,{~~~~}
	\end{eqnarray}
	and
	\begin{eqnarray}
		\label{eq:FermiFunctions}
		F_0(Z,E_e) &=& \frac{{g}_{-1}(E_e) {g}_{-1}(E_e) + {f}_{1}(E_e) {f}_{1}(E_e)}
		{j_0(p_eR)~j_0(p_eR)}, \nonumber\\
		F_1(Z,E_e) &=& \frac{{g}_{-2}(E_e) {g}_{-2}(E_e) + {f}_{2}(E_e) {f}_{2}(E_e)}
		{j_1(p_eR)~j_1(p_eR)}.~
	\end{eqnarray}
	Here, $g_{\kappa}(E_e) \equiv g_{\kappa}(E_e,R)$ and $f_{\kappa}(E_e) \equiv f_{\kappa}(E_e,R)$.
	In the limit the Coulomb interaction is switched-off $F_0(Z,E_e) =1$ and $F_1(Z,E_e) =1$. Different approximation schemes, for the electron wave functions, are indicated in the Fermi functions as $F_{k-1}^{\rm I}(Z,E_e)$, where I = A, B, C or D.

	\section{Exchange correction}
	
	\label{sec:ExchnageCorrection}

	Besides the screening correction, another prominent atomic correction, significant for low Q-value $\beta$ transitions, is the exchange effect. It arises from creating a $\beta$ electron in a bound orbital of the final atom corresponding to one occupied in the initial atom. An atomic electron from the bound orbital simultaneously makes a transition to a continuum state of the final atom \cite{HarstonPRA1992, PyperPRSLA1998}.
	
	We have extended the exchange effect formalism for allowed $\beta$ transitions, presented in \cite{HarstonPRA1992,PyperPRSLA1998}, for the unique first  forbidden $\beta$ transitions. We found that two components of the exchange correction independently modify the spectra associated with the emission of electrons in the $s_{1/2}$ and $p_{3/2}$ states,    
	\begin{eqnarray}
		\frac{d {\mit\Gamma}^{s_{1/2}}}{dE_e}&\Rightarrow&\frac{d {\mit\Gamma}^{s_{1/2}}}{dE_e}\times(1+\eta^{\text{T}}_1(E_e))\nonumber\\
		\frac{d {\mit\Gamma}^{p_{3/2}}}{dE_e}&\Rightarrow&\frac{d {\mit\Gamma}^{p_{3/2}}}{dE_e}\times(1+\eta^{\text{T}}_2(E_e))
	\end{eqnarray}  
	
	The result is consistent with the one obtained in \cite{Haselschwardt-PRC2020}, for the unique first forbidden $\beta$-decay of $^{85}$Kr. The total exchange correction for each partial spectrum is given by, 
	\begin{eqnarray}
		\label{eq:TotalEchangeCorrection}
		\eta_k^T(E_e)&=&f_{k}(2T_{-k}+T_{-k}^2)+(1-f_{k})(2T_{+k}+T_{+k}^2)\nonumber\\
		&=&\eta_{-k}(E_e)+\eta_{+k}(E_e) 
	\end{eqnarray}
	where $k=\left|\kappa\right|$ can take the values $1$ or $2$. Here,
	\begin{eqnarray}
		f_k=\frac{g'^2_{-k}(E_e,R)}{g'^2_{-k}(E_e,R)+f'^2_{+k}(E_e,R)},
	\end{eqnarray}
	and the dimensionless quantities $T_{\kappa}$ depend on the overlaps between the bound states of the initial atom and the continuum states of the final atom with energy $E_e$,
	\begin{eqnarray}
		\label{eq:TnPlusQuantities}
		T_{\kappa}=\sum_{(n\kappa)'}T_{n\kappa}=-\sum_{(n\kappa)'}\frac{\braket{\psi'_{E_e\kappa}}{\psi_{n\kappa}}}{\braket{\psi'_{n\kappa}}{\psi_{n\kappa}}}\frac{g'_{n,\kappa}(R)}{g'_{\kappa}(E_e,R)},
	\end{eqnarray}
	for electrons in $s_{1/2}$ ($\kappa=-1$) and $p_{3/2}$ ($\kappa=-2$) states, and
	\begin{eqnarray}
		\label{eq:TnMinusQuantities}
		T_{\kappa}=\sum_{(n\kappa)'}T_{n\kappa}=-\sum_{(n\kappa)'}\frac{\braket{\psi'_{E_e\kappa}}{\psi_{n\kappa}}}{\braket{\psi'_{n\kappa}}{\psi_{n\kappa}}}\frac{f'_{n,\kappa}(R)}{f'_{\kappa}(E_e,R)},
	\end{eqnarray}
	for electrons in $p_{1/2}$ ($\kappa=+1$) and $d_{3/2}$ ($\kappa=+2$) states. All primed continuum and bound states refer to the final atom. The sums are running over all occupied orbitals of the final atom, which, in the sudden approximation, correspond to the parent electronic configuration.

	Taking into account the sums inside the $T_{\kappa}$ quantities, we can write
	\begin{equation}
		\label{eq:PartialExchnageSOrbitals}
		\eta_{\kappa}(E_e)=\sum_{n}\eta_{n\kappa}+f_{\left|\kappa\right|}\sum\limits_{\substack{n,m \\ n\neq m}}T_{n\kappa}T_{m\kappa}
	\end{equation}  
	for negative values of $\kappa$, and 
	\begin{equation}
		\eta_{\kappa}(E_e)=\sum_{n}\eta_{n\kappa}+(1-f_\kappa)\sum\limits_{\substack{n,m \\ n\neq m}}T_{n\kappa}T_{m\kappa}
	\end{equation}
	for positive values of $\kappa$. In this way, we can define the partial exchange correction contributions, $\eta_{n\kappa}$, given by 
	\begin{equation}
		\label{eq:PartialExchnageMinusK}
		\eta_{n\kappa}=f_{\left|\kappa\right|}(2T_{n\kappa}+T^2_{n\kappa})
	\end{equation}
	for negative values of $\kappa$, and 
	\begin{equation}
		\label{eq:PartialExchnagePlusK}
		\eta_{n\kappa}=(1-f_{\kappa})(2T_{n\kappa}+T^2_{n\kappa}),
	\end{equation}
	for positive values of $\kappa$.

	\begin{figure*}
		\includegraphics[width=0.88\textwidth]{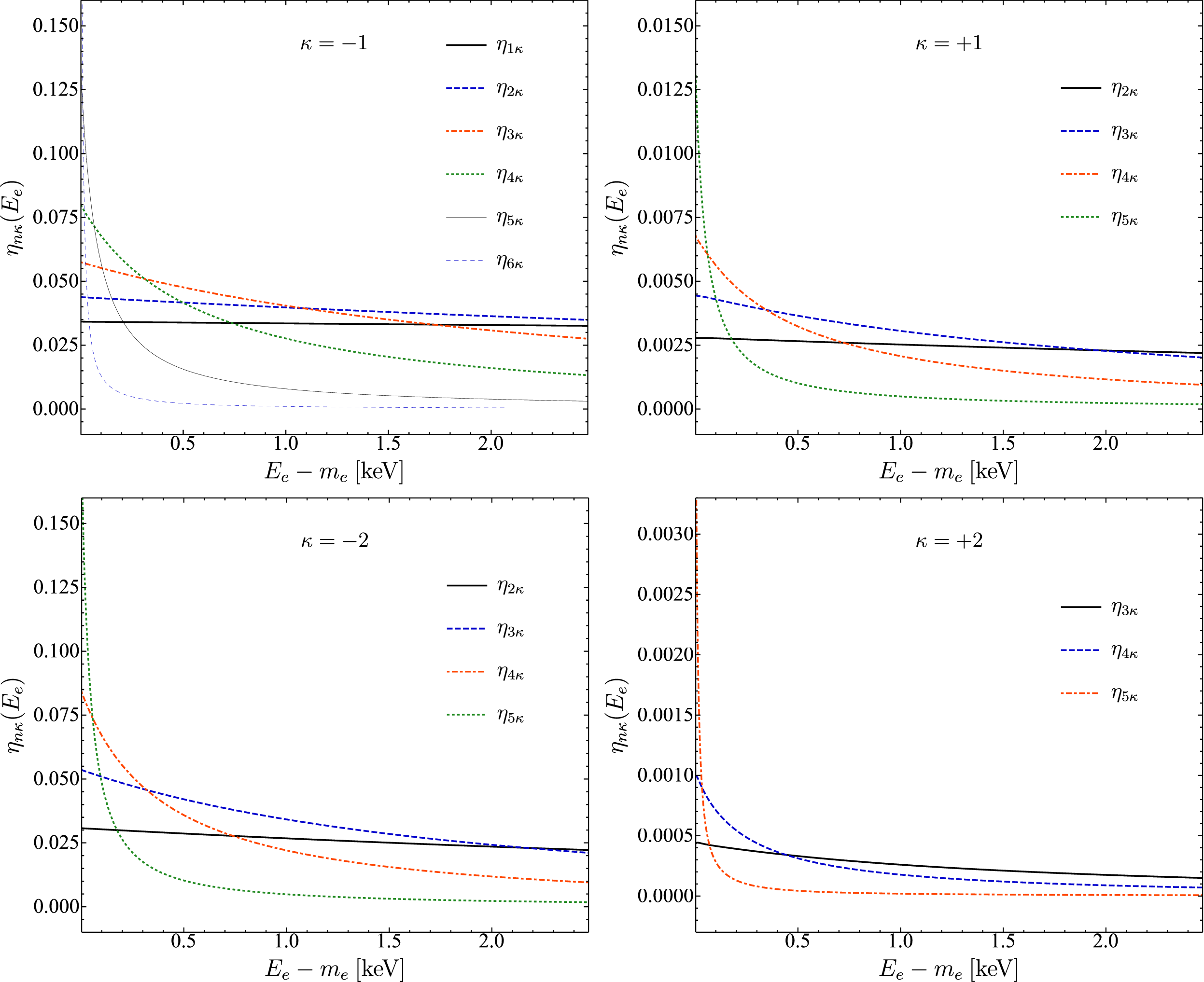}
		\caption{All partial exchange corrections for the decay of $^{187}$Re, i.e., Eqs.~(\ref{eq:PartialExchnageMinusK}) and (\ref{eq:PartialExchnagePlusK}). The results are presented for atomic electrons in $s_{1/2}$ ($\kappa=-1$), $p_{3/2}$ ($\kappa=-2$), $p_{1/2}$ ($\kappa=+1$) and $d_{3/2}$ ($\kappa=+2$) states.\label{fig:PartialExchangeContributions}} 
	\end{figure*}
	
	\begin{figure}[h]
		\includegraphics[width=0.48\textwidth]{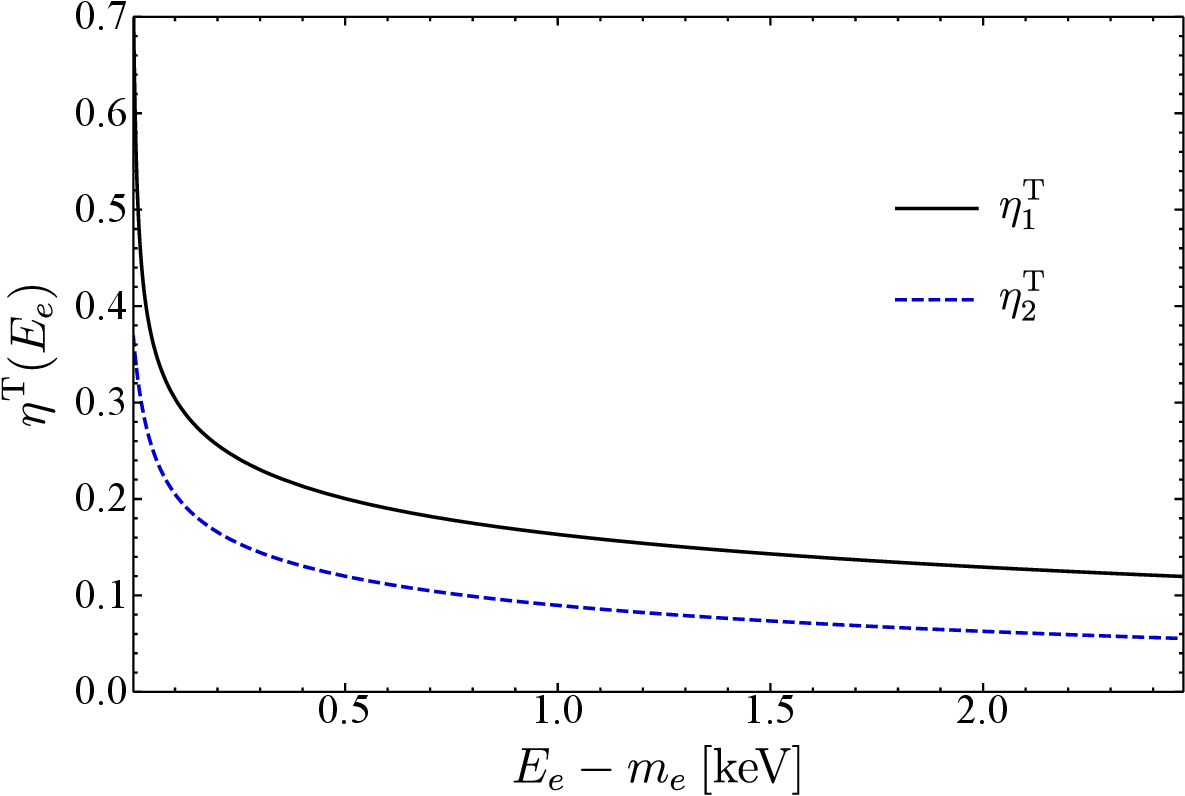}
		\caption{The total exchange correction for electrons emitted in $s_{1/2}$ wave state, $\eta^{\text{T}}_1(E_e)$ (solid black line), and for electrons emitted in $p_{3/2}$ wave state, $\eta^{\text{T}}_2(E_e)$ (dashed blue line).     \label{fig:TotalExchange}} 
	\end{figure}

	The essential quantities in computing the exchange correction are the overlaps between the continuum state electron wave function, with energy $E_e$, in the final atom and the bound orbitals electron wave functions in the initial atom, i.e., $\braket{\psi'_{E_es}}{\psi_{ns}}$. This overlap is given explicitly by
	\begin{eqnarray}
		\braket{\psi'_{E_e\kappa}}{\psi_{n\kappa}}&=&\int_{0}^{\infty}r^2g'_{\kappa}(E_e,r)g_{n,\kappa}(r)dr \nonumber\\
		&+&\int_{0}^{\infty}r^2f'_{\kappa}(E_e,r)f_{n,\kappa}(r)dr.
	\end{eqnarray}
	and its numerical calculation requires good knowledge of the continuum wave function over a wide region of space, from the nuclear center to where the bound wave function for the initial atom ends.
	
	In the calculation of the exchange correction, the most important condition is to ensure that the final-state electron continuum wave function is orthogonal to the wave functions of the final-state bound orbitals, i.e., $\braket{\psi'_{E_e\kappa}}{\psi'_{n\kappa}}=0$. They are eigenfunctions of the same Hamiltonian, so they must be orthogonal. A non-zero value of $\braket{\psi'_{E_e\kappa}}{\psi'_{n\kappa}}$ may lead to significant error in $\braket{\psi'_{E_e\kappa}}{\psi_{n\kappa}}$ \cite{HarstonPRA1992}. The implications in the exchange correction of non-orthogonal continuum and bound final-states wave functions were discussed in detail in \cite{NitescuPRC2023}. Our chose of a modified DHFS self-consistent method for bound states and of the potential presented in Eq.~(\ref{eq:PotentialSchemeD}) for continuum states, imposes automatically the orthogonality condition, $\braket{\psi'_{E_e\kappa}}{\psi'_{n\kappa}}=0$, without any further numerically expensive orthogonalization procedures.

	In the case of $^{187}$Re, there is an abundance of possible switches between the $\beta$ emitted electrons and the bound electrons from the atomic cloud. For the emission of $s_{1/2}$-state electrons, ten orbitals contribute to $\eta_1^{\rm T}$, and for the dominant emission in $p_{3/2}$-state, seven orbitals contribute to $\eta_2^{\rm T}$. In Fig.~\ref{fig:PartialExchangeContributions}, we present all partial exchange corrections for the $\beta$ decay of $^{187}$Re, i.e., Eqs.~(\ref{eq:PartialExchnageMinusK}) and (\ref{eq:PartialExchnagePlusK}), as functions of the kinetic energy of the emitted electrons, i.e., $E_e-m_e$. We note that the contributions from $p_{1/2}$ (top right panel) and $d_{3/2}$ (bottom right panel) states are much smaller than the ones from $s_{1/2}$ (top left panel)  and $p_{3/2}$ (bottom left panel) states because they are associated with the small components of the Fermi functions $f_1(E_e)$ and $f_2(E_e)$, respectively (see Eq.~(\ref{eq:FermiFunctions})).  They can not be neglected for a precise calculation.

	The total exchange correction for $s_{1/2}$ ($p_{3/2}$) electron spectrum of rhenium $\beta$ decay is depicted with a solid (dashed) line in Fig.~\ref{fig:TotalExchange} as a function of the kinetic energy of the emitted electron. We can see that the exchange corrections decrease with the increasing energy of the emitted electron, starting from around $73\%$ for $s_{1/2}$ electrons and $37\%$ for $p_{3/2}$ electrons, at $2$ eV kinetic energy. The exchange correction remains significant even in the $Q$-value region, around $12\%$ for $s_{1/2}$ electrons and $5.5\%$ for $p_{3/2}$ electrons. Considering the large amplitude and the shape of the exchange correction, we expect that it will induce considerable modifications in both the decay rate and spectrum shape of the $\beta$-decay of rhenium. Those modifications are discussed in what follows.

	\section{Results and Discussions}

	In Table~\ref{tab:DecayRates}, we present the partial decay rates,  $\Gamma^{s_{1/2}}$ (second column) and $\Gamma^{p_{3/2}}$ (fourth column), from which we excluded the square matrix element, $B$. The partial decay rates, presented in MeV, have been obtained from the integration over the full energy range of Eq.~(\ref{eq.sp}). In each integration, we have used different approximation schemes (indicated in the first column) displayed in Section~\ref{sec:ApproximationSchemes}. The last line corresponds to the approximation scheme D, where we also included the atomic exchange correction, presented in Section~\ref{sec:ExchnageCorrection}. In what follows, this combination is indicated as $\rm D+ex$. We consider the approximation scheme A, used in our previous investigation, as a reference for a percent deviation with the other approximation schemes. The deviations are presented in the third (fifth) column for the partial decay rate $\Gamma^{s_{1/2}}$ ($\Gamma^{p_{3/2}}$). Using the experimental half-live, $4.33\times 10^{10}$ y \cite{BasuniaNDS2009}, we also present our prediction for the squared matrix elements in the last column of Table~\ref{tab:DecayRates}. 
	
	\begin{table}[h]
		\caption{\label{tab:DecayRates} The partial decay rates for $^{187}$Re, excluding the squared matrix element, associated with the emission of electrons in $s_{1/2}$ wave state (second colum) and $p_{3/2}$ wave state (forth column). The approximation scheme for the electron wave functions is indicated in the first column. In the last line, the addition of the exchange correction over the scheme D is indicated as $\rm D+ex$. In the third and fifth column, we present the decay rate percent deviation between the scheme A and the other schemes, $\delta^{s_{1/2}}=100(\Gamma_{\rm{A}}^{s_{1/2}}-\Gamma_{\rm{X}}^{s_{1/2}})/\Gamma_{\rm{X}}^{s_{1/2}}$, and $\delta^{p_{3/2}}=100(\Gamma_{\rm{A}}^{p_{3/2}}-\Gamma_{\rm{X}}^{p_{3/2}})/\Gamma_{\rm{X}}^{p_{3/2}}$, respectively, where X can be B, C, D or $\rm D+ex$. The last column presents the experimental squared matrix elements.  
		}
		\begin{ruledtabular}
			\begin{tabular}{cccccc}
				\textrm{ w. f.}&
				$\frac{10^{41}}{B}\times\Gamma^{s_{1/2}}$ &
				$\delta^{s_{1/2}}$&
				$\frac{10^{37}}{B}\times\Gamma^{p_{3/2}}$ &
				$\delta^{p_{3/2}}$&
				$B\times10^4$\\
				& [MeV] &\% & [MeV] &\% &\\
				\colrule
				\textrm{A} & 9.30 &-&  9.19 & -& 3.63 \\
				\textrm{B} & 8.33 &-10.41&  8.92 &  -2.95& 3.74\\
				\textrm{C} & 7.88 &-15.23& 8.88  &-3.35&3.76\\
				\textrm{D} & 7.58 & -18.48& 6.98 &-24.02&4.78\\
				\textrm{D+ex} & 9.46 & 1.75& 7.92&-13.84&4.22\\
			\end{tabular}
		\end{ruledtabular}
	\end{table}

	As a general result, in all the approximation schemes, the ratios between $p_{3/2}$-state electron emission channel and the $s_{1/2}$-state electron emission one are always around $10^4$, so different corrections do not change this particular feature of the rhenium decay. Still, in comparison with scheme A, there are considerable differences in the decay rates due to the screening and exchange corrections. There is a $24\%$ ($18.5\%$) decrease in the partial decay rate $\Gamma^{p_{3/2}}$ ($\Gamma^{s_{1/2}}$) due to screening correction alone (scheme D). If we also include the exchange correction (scheme $\rm D+ex$), the decay rate $\Gamma^{s_{1/2}}$ returns close to the one from scheme A (just a $1.7\%$ increase). It is not the case of $\Gamma^{p_{3/2}}$, which in scheme $\rm D+ex$, is $13.8\%$ lower than the value from scheme A.

	\begin{figure}[h]
		\includegraphics[width=0.48\textwidth]{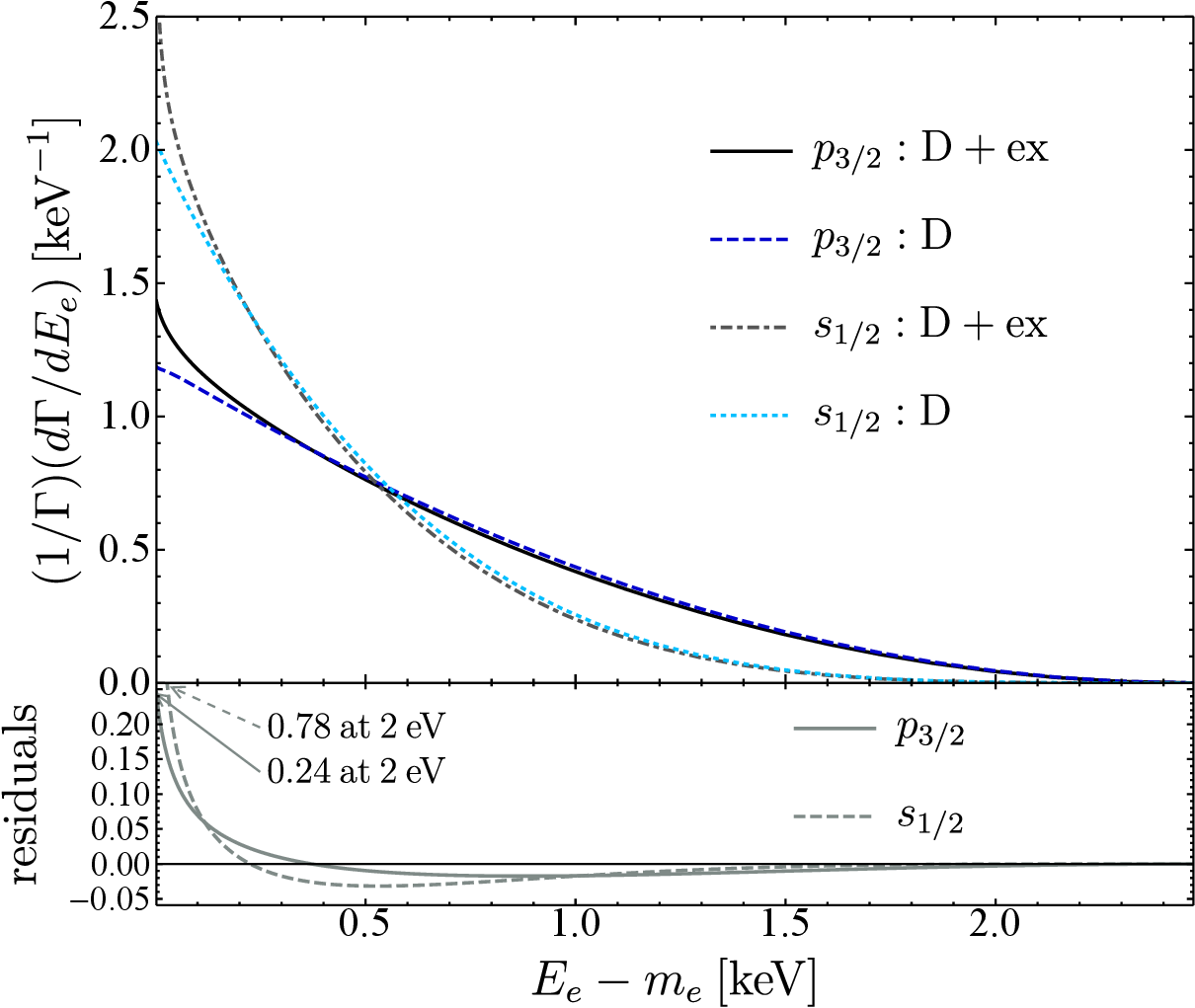}
		\caption{The single electron differential decay rate normalized to the particular decay rate ($\Gamma^{s_{1/2}}$ and $\Gamma^{p_{3/2}}$) for the emission of $s_{1/2}$ and $p_{3/2}$ electrons as functions of the electron kinetic energy $E_e-m_e$ for $\beta$-decay of $^{187}$Re. We indicate with D the approximation scheme D and with ex the atomic exchange correction. The lower portion of the figure gives the difference between the spectra with and without the exchange correction for $s_{1/2}$ wave state electrons with a dashed line and $p_{3/2}$ wave state electrons with a solid line. Spectra are normalized over the full energy range.     \label{fig:Spectra}} 
	\end{figure}
	
	The most striking modification due to the exchange correction is the change in the spectrum shape of the rhenium decay. In Fig.~\ref{fig:Spectra}, we plot the normalized single electron spectrum for both $s_{1/2}$- and $p_{3/2}$-state emissions, using the approximation schemes D and  $\rm D+ex$. We mention that there are no considerable shape differences between approximation schemes A, B, C, and D.  In the lower portion of Fig.~\ref{fig:Spectra}, we also present the residuals between spectra. Although the dominant $p_{3/2}$ spectrum is not as strongly influenced by the exchange correction as the $s_{1/2}$ spectrum, the modification is big enough to change the shape of the total spectrum of the decay.

\begin{figure}[h]
		\includegraphics[width=0.48\textwidth]{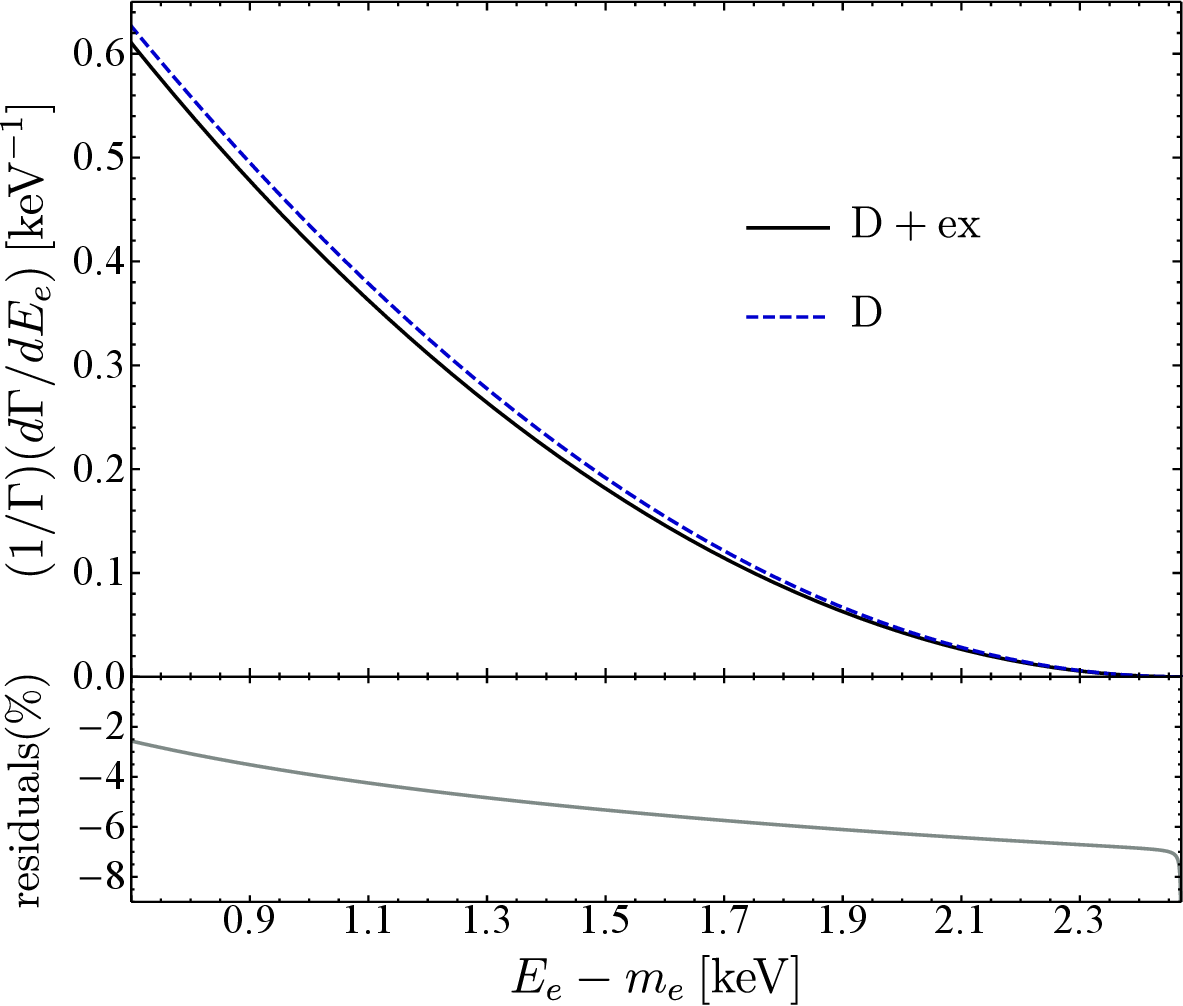}
		\caption{The differential decay rate normalized to the total decay rate ($\Gamma$) as function of the electron kinetic energy $E_e-m_e$ for $\beta$ decay of $^{187}$Re. We indicate with D the approximation scheme D and with ex the atomic exchange correction. The lower portion of the figure gives the percentage residuals between the spectrum with and without the exchange correction. Spectra are normalized over the full energy range.  \label{fig:TotalSpectra}} 
	\end{figure}

	We present the total spectrum of rhenium $\beta$ decay in Fig.~\ref{fig:TotalSpectra} with the same convention as in Fig.~\ref{fig:Spectra}, but now the electron kinetic energy takes values from $700$ keV up to the $Q$-value, and the residuals are in percentages. Note that the spectra are normalized to unity over the full range of the kinetic energy. One can see a different shape of the total electron spectrum of the rhenium decay when the exchange correction is switched on. The considerable spectrum shape modification induced by the exchange correction indicates that this correction should be included in the neutrino mass investigations from rhenium decay, particularly, and other low $Q$-value $\beta$ transitions, in general.
	
	To evaluate the deviation from an allowed spectrum, we write the $\beta$ spectrum of $^{187}$Re as,
	\begin{eqnarray}
		\frac{d {\Gamma}}{dE_e}=\frac{G_F^2 V_{ud}^2}{2 \pi^3}
		B p_e E_eF^{\rm I}_{0}(Z,E_e) (E_0 - E_e)^2 A_F^{\rm I}
	\end{eqnarray}
	where, for the moment, the neutrino masses are neglected and I = A, B, C, or D. The shape factor $ A_F^{\rm I}=1$ for the allowed transitions, but for unique first forbidden,
	\begin{eqnarray}
		\label{eq:ShapeFactorABCD}
		A_F^{\rm I}=\frac{R^2}{3}\left[p_e\frac{F^{\rm I}_1(Z,E_e)}{F^{\rm I}_0(Z,E_e)}+(E_0-E_e)^2\right],
	\end{eqnarray}
	where we did not include the exchange with bound electrons. If we want to take into account this effect, the shape factor becomes,
	\begin{eqnarray}
		\label{eq:ShapeFactorExchange}
		A_F^{\rm{D+ex}}=\frac{R^2}{3}\biggl\{&p_e&\frac{F^{\rm D}_1(Z,E_e)}{F^{\rm D}_0(Z,E_e)}(1+\eta^{\text{T}}_2(E_e)) \nonumber \\
		&+&(E_0-E_e)^2(1+\eta^{\text{T}}_1(E_e))\biggl\}.
	\end{eqnarray}
	
\begin{figure}[h]
		\includegraphics[width=0.48\textwidth]{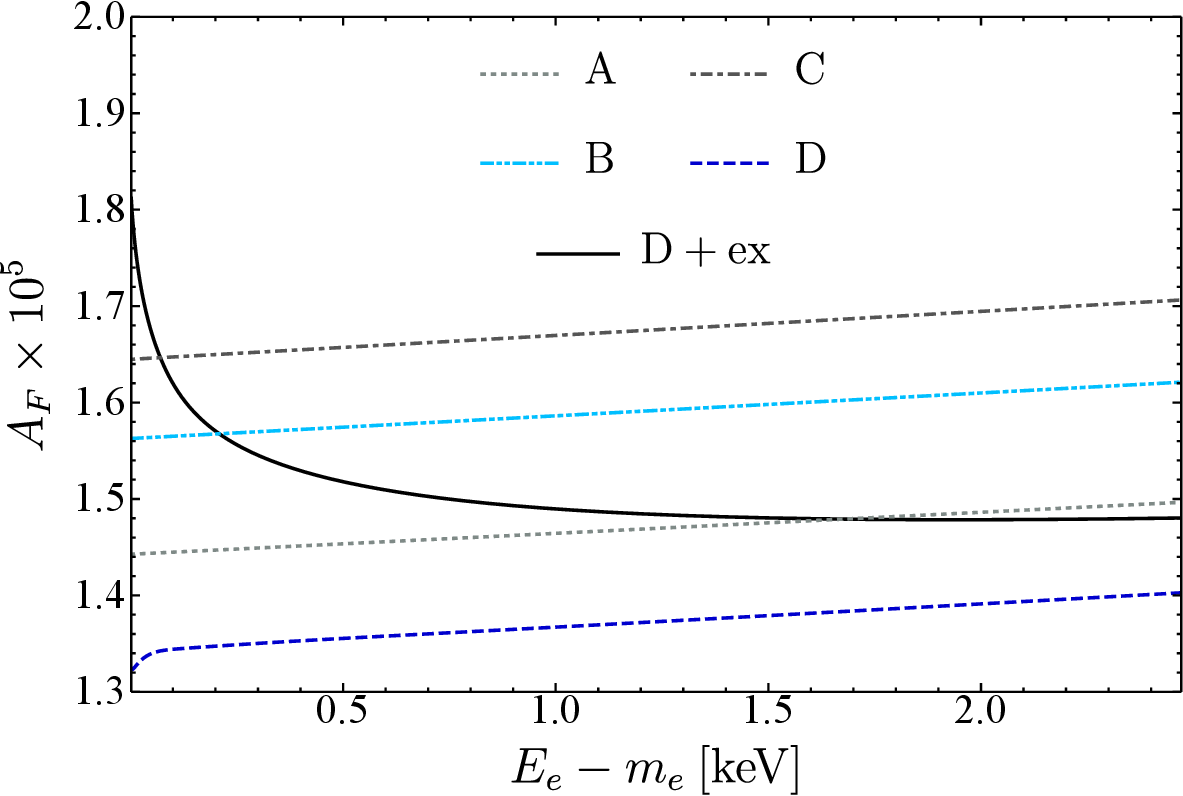}
		\caption{The shape factor $A_F^{\rm I}$, i.e. Eq.~(\ref{eq:ShapeFactorABCD}), for different approximation schemes, I = A, B, C and D, and the shape factor $A_F^{\rm D+ex}$, i.e. Eq.~(\ref{eq:ShapeFactorExchange}).   \label{fig:ShapeFactors}} 
	\end{figure}

	We can see from Fig.~\ref{fig:ShapeFactors}, that all $A_F^{\rm I}$ are linear increase with energy in the experimentally available range, from $700$ eV to $Q$-value. We reproduce perfectly $A_F^{\rm I}$ if we consider the following model fit,
	\begin{eqnarray}
		A_F^{\rm I}=a^{\rm I}\left(1+b_1T_e+b_2T_e^2\right)
	\end{eqnarray}
	with $T_e=E_e-m_e$ in eV. The best fit parameters are $b_1=1.50\times10^{-5}$ eV$^{-1}$ and $b_2=4.82\times10^{-11}$ eV$^{-2}$ The parameters $a^{\rm I}$ simply scale through different approximations of the Fermi functions (for example, $a^{\rm A}=1.44\times10^{-5}$), but they do not change the linear increasing behavior of $A_F^{\rm I}$. 
	In contrast, $A_F^{\rm{D+ex}}$ decreases with energy, and its shape is more complex due to the exchange correction. The best fit in the range from $700$ eV to the $Q$-value is obtained with the following assumption,
	\begin{eqnarray}
		A_F^{\rm{D+ex}}= a^{\rm{D+ex}}\left(\frac{b_{-1}}{T_e}+1+b_1T_e+b_2T_e^2\right),
	\end{eqnarray}
	where we found the best fit parameters $b_{-1}=19.95$ eV, $b_1=-6.80\times10^{-6}$ eV$^{-1}$ and $b_2=3.05\times10^{-9}$ eV$^{-2}$ with $a^{\rm{D+ex}}=1.46\times10^{-5}$. Important to notice is the sign change of $b_1$ when the exchange correction is included.  The choice of a fit model with terms proportional to $T_e^{-1}$ is based on the experimental shape factors that can also involve such terms \cite{MougeotPRC2015}.

	The best upper limit on effective neutrino mass, $m_\beta \leqslant 0.8$ eV, recently reported by the KATRIN experiment \cite{KATRIN-N2022}, holds in the degenerate neutrino mass region, i.e., $m_1\simeq m_2\simeq m_3\simeq m_\beta=\sum_{k=1}^{3}\left|U_{ek}\right|^2 m_k$. So, we replace all the neutrino masses $m_k$ ($k=1,2,3$) with the effective neutrino mass $m_\beta$. For the following discussion we consider just the approximation scheme D for the electron wave functions and we also include the atomic exchange correction.  
	
	\begin{figure}[h]
		\includegraphics[width=0.48\textwidth]{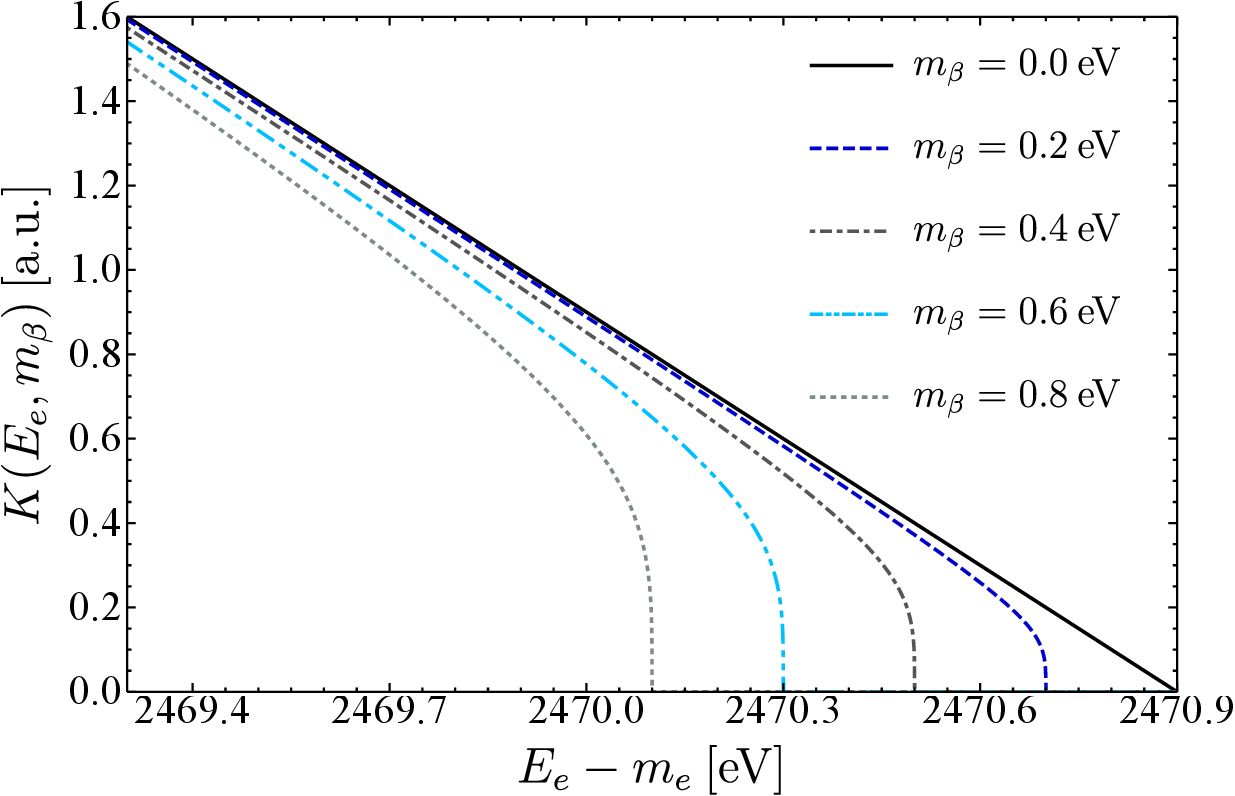}
		\caption{The Kurie plots in arbitrary units (a.u.) for the $\beta$ decay of $^{187}$Re with different values of the effective neutrino mass: $m_\beta=0.0$, $0.2$, $0.4$, $0.6$, and $0.8$ eV. The Q-value considered is $2470.9$ eV \cite{Qvalue187Re}.    \label{fig:KuriePlots
		}} 
	\end{figure}
	
	The Kurie functions for the unique first forbidden transitions is given by
	\begin{eqnarray}
		K(E_e,m_\beta)&=&\sqrt{\frac{d\Gamma/dE_e}{p_eE_e(p_eR)^2F_1(Z,E_e)(1+\eta^{\text{T}}_2(E_e))}}\nonumber\\
		&=&G_F V_{ud}\sqrt{\frac{B}{6\pi^3}}(E_0-E_e)\sqrt[4]{1-\frac{m^2_\beta}{(E_0-E_e)^2}} \nonumber\\
		&\times&\left[1+\frac{p_\nu}{p_e}\frac{F_0(Z,E_e)}{F_1(Z,E_e)}\frac{(1+\eta^{\text{T}}_1(E_e))}{(1+\eta^{\text{T}}_2(E_e))}\right]^{1/2}.
	\end{eqnarray}
	In the case of $m_\beta=0$, the Kurie plot, which is a plot of $K(E_e,m_\beta)$ versus $E_e$, is ensured to be linear by keeping the exchange correction for $p_{3/2}$ wave state electrons in the denominator of its definition. The term that can change the linear behavior is the last term in squared brackets, but its deviation from unity is very small. This deviation is below $6\times10^{-5}$ after $1000$ eV kinetic energy, and in the region of interest from $2300$ eV to the $Q$-value, the deviation is even lower under $10^{-6}$.

	\section{Conclusions}
	
	The distortion in the endpoint measurements of low $Q$-value $\beta$ decay spectra is a direct tool to measure the values of neutrino masses. Current experiments are based on the ground-state to ground-state $\beta$ transitions, and for next-generation investigations, some suitable candidates could be the ultra-low $Q$ value (under $1$ keV) ground-state-to-excited-state $\beta$ transitions. With the increasing interest in measuring neutrino masses from $\beta$ decays, accurate theoretical spectra predictions should be provided. However, their description can be challenging considering the $\beta$ transition classification and the plurality of atomic effects in the low-energy region. 
	
	We investigate the ground-state to ground-state unique first forbidden $\beta$ transition from $^{187}$Re($5/2^+$) to $^{187}$Os($1/2^-$), considering all relevant corrections to its spectrum and decay rate. In addition to our previous paper, we incorporate corrections for finite nuclear size, diffuse nuclear surface, screening, and exchange in the rhenium $\beta$ decay model. The last two effects are calculated using the self-consistent Dirac-Hartree-Fock-Slater description for the atomic bound electrons of the final atom. As the rhenium $\beta$ emission involves a mixture of $s_{1/2}$-state and $p_{3/2}$-state electrons, our exchange correction calculation accounts for all possible contributions from exchanges with $s_{1/2}$, $p_{3/2}$, $p_{1/2}$, and $d_{3/2}$ bound orbitals. Our findings reveal significant alterations in the partial decay rates for both $s_{1/2}$- and $p_{3/2}$-state emission channels due to the screening and exchange effects while maintaining the experimentally confirmed dominance of $p_{3/2}$-state emission. 
	
	The key feature of this paper is that, in addition to changing the partial decay rates, the exchange correction modifies the shape of the total electron spectrum for rhenium $\beta$ decay. By studying the deviations from an allowed spectrum, we demonstrate that calculations with and without the exchange effect lead to entirely different shape factors, transitioning from an increasing linear behavior to a decreasing quadratic one. We present the best-fit parameters for both cases. The extent of this shape modification is so pronounced that we need to include the exchange correction in the definition of the Kurie plot to maintain its linearity in the case of zero effective neutrino mass. We also demonstrate how different effective neutrino masses affect the Kurie plots near the endpoint of the $\beta$ decay of $^{187}$Re. Our conclusion underscores the importance of considering atomic effects, especially the exchange effect, in current and future neutrino mass scale investigations from $\beta$ decays.
	
	\section*{ACKNOWLEDGMENTS}

	F.\v{S}. acknowledges support by the Slovak Research and Development Agency under Contract No. APVV-22-0413, and by the Ministry of Education, Youth and Sports of the Czechia under the INAFYM Grant No. CZ.02.1.01/0.0/0.0/16\_019/0000766. O.N. acknowledges support support of the
	Romanian Ministry of Research, Innovation, and Digitalization through Project CNCS-UEFISCDI No. 99/2021 within PN-III-P4-ID-PCE-2020-237 and the project CNCS-UEFISCDI No. TE12/2021 within PN-III-P1-1.1-TE-2021-0343.

	The figures for this article have been created using the SciDraw scientific figure preparation system \cite{SciDraw}.
	\bibliography{myAwesomeBiblio}

\providecommand{\noopsort}[1]{}\providecommand{\singleletter}[1]{#1}%
\begin{thebibliography}{46}%
\makeatletter
\providecommand \@ifxundefined [1]{%
 \@ifx{#1\undefined}
}%
\providecommand \@ifnum [1]{%
 \ifnum #1\expandafter \@firstoftwo
 \else \expandafter \@secondoftwo
 \fi
}%
\providecommand \@ifx [1]{%
 \ifx #1\expandafter \@firstoftwo
 \else \expandafter \@secondoftwo
 \fi
}%
\providecommand \natexlab [1]{#1}%
\providecommand \enquote  [1]{``#1''}%
\providecommand \bibnamefont  [1]{#1}%
\providecommand \bibfnamefont [1]{#1}%
\providecommand \citenamefont [1]{#1}%
\providecommand \href@noop [0]{\@secondoftwo}%
\providecommand \href [0]{\begingroup \@sanitize@url \@href}%
\providecommand \@href[1]{\@@startlink{#1}\@@href}%
\providecommand \@@href[1]{\endgroup#1\@@endlink}%
\providecommand \@sanitize@url [0]{\catcode `\\12\catcode `\$12\catcode
  `\&12\catcode `\#12\catcode `\^12\catcode `\_12\catcode `\%12\relax}%
\providecommand \@@startlink[1]{}%
\providecommand \@@endlink[0]{}%
\providecommand \url  [0]{\begingroup\@sanitize@url \@url }%
\providecommand \@url [1]{\endgroup\@href {#1}{\urlprefix }}%
\providecommand \urlprefix  [0]{URL }%
\providecommand \Eprint [0]{\href }%
\providecommand \doibase [0]{https://doi.org/}%
\providecommand \selectlanguage [0]{\@gobble}%
\providecommand \bibinfo  [0]{\@secondoftwo}%
\providecommand \bibfield  [0]{\@secondoftwo}%
\providecommand \translation [1]{[#1]}%
\providecommand \BibitemOpen [0]{}%
\providecommand \bibitemStop [0]{}%
\providecommand \bibitemNoStop [0]{.\EOS\space}%
\providecommand \EOS [0]{\spacefactor3000\relax}%
\providecommand \BibitemShut  [1]{\csname bibitem#1\endcsname}%
\let\auto@bib@innerbib\@empty
\bibitem [{\citenamefont {{Aghanim, N.}}\ \emph {et~al.}(2020)\citenamefont
  {{Aghanim, N.}} \emph {et~al.}}]{CosmologyNeutrinoMassAA2022}%
  \BibitemOpen
  \bibfield  {author} {\bibinfo {author} {\bibnamefont {{Aghanim, N.}}} \emph
  {et~al.} (\bibinfo {collaboration} {Planck Collaboration}),\ }\bibfield
  {title} {\bibinfo {title} {Planck 2018 results - {VI}. {C}osmological
  parameters},\ }\href {https://doi.org/10.1051/0004-6361/201833910} {\bibfield
   {journal} {\bibinfo  {journal} {A\&A}\ }\textbf {\bibinfo {volume} {641}},\
  \bibinfo {pages} {A6} (\bibinfo {year} {2020})}\BibitemShut {NoStop}%
\bibitem [{\citenamefont {Avignone}\ \emph {et~al.}(2008)\citenamefont
  {Avignone}, \citenamefont {Elliott},\ and\ \citenamefont
  {Engel}}]{AvignoneRMP2008}%
  \BibitemOpen
  \bibfield  {author} {\bibinfo {author} {\bibfnamefont {F.~T.}\ \bibnamefont
  {Avignone}}, \bibinfo {author} {\bibfnamefont {S.~R.}\ \bibnamefont
  {Elliott}},\ and\ \bibinfo {author} {\bibfnamefont {J.}~\bibnamefont
  {Engel}},\ }\bibfield  {title} {\bibinfo {title} {Double beta decay, majorana
  neutrinos, and neutrino mass},\ }\href
  {https://doi.org/10.1103/RevModPhys.80.481} {\bibfield  {journal} {\bibinfo
  {journal} {Rev. Mod. Phys.}\ }\textbf {\bibinfo {volume} {80}},\ \bibinfo
  {pages} {481} (\bibinfo {year} {2008})}\BibitemShut {NoStop}%
\bibitem [{\citenamefont {Ejiri}\ \emph {et~al.}(2019)\citenamefont {Ejiri},
  \citenamefont {Suhonen},\ and\ \citenamefont {Zuber}}]{EjiriPR2019}%
  \BibitemOpen
  \bibfield  {author} {\bibinfo {author} {\bibfnamefont {H.}~\bibnamefont
  {Ejiri}}, \bibinfo {author} {\bibfnamefont {J.}~\bibnamefont {Suhonen}},\
  and\ \bibinfo {author} {\bibfnamefont {K.}~\bibnamefont {Zuber}},\ }\bibfield
   {title} {\bibinfo {title} {Neutrino-nuclear responses for astro-neutrinos,
  single beta decays and double beta decays},\ }\href
  {https://doi.org/https://doi.org/10.1016/j.physrep.2018.12.001} {\bibfield
  {journal} {\bibinfo  {journal} {Physics Reports}\ }\textbf {\bibinfo {volume}
  {797}},\ \bibinfo {pages} {1} (\bibinfo {year} {2019})}\BibitemShut {NoStop}%
\bibitem [{\citenamefont {{\v{S}}imkovic}(2021)}]{SimkovicPU2021}%
  \BibitemOpen
  \bibfield  {author} {\bibinfo {author} {\bibfnamefont {F.}~\bibnamefont
  {{\v{S}}imkovic}},\ }\bibfield  {title} {\bibinfo {title} {Neutrino masses
  and interactions and neutrino experiments in the laboratory},\ }\href
  {https://doi.org/10.3367/ufne.2021.08.039036} {\bibfield  {journal} {\bibinfo
   {journal} {Physics-Uspekhi}\ }\textbf {\bibinfo {volume} {64}},\ \bibinfo
  {pages} {1238} (\bibinfo {year} {2021})}\BibitemShut {NoStop}%
\bibitem [{\citenamefont {Ferri}\ \emph {et~al.}(2015)\citenamefont {Ferri},
  \citenamefont {Bagliani}, \citenamefont {Biasotti}, \citenamefont {Ceruti},
  \citenamefont {Corsini}, \citenamefont {Faverzani}, \citenamefont {Gatti},
  \citenamefont {Giachero}, \citenamefont {Gotti}, \citenamefont {Kilbourne},
  \citenamefont {Kling}, \citenamefont {Maino}, \citenamefont {Manfrinetti},
  \citenamefont {Nucciotti}, \citenamefont {Pessina}, \citenamefont
  {Pizzigoni}, \citenamefont {Gomes},\ and\ \citenamefont
  {Sisti}}]{FerriPP2015}%
  \BibitemOpen
  \bibfield  {author} {\bibinfo {author} {\bibfnamefont {E.}~\bibnamefont
  {Ferri}}, \bibinfo {author} {\bibfnamefont {D.}~\bibnamefont {Bagliani}},
  \bibinfo {author} {\bibfnamefont {M.}~\bibnamefont {Biasotti}}, \bibinfo
  {author} {\bibfnamefont {G.}~\bibnamefont {Ceruti}}, \bibinfo {author}
  {\bibfnamefont {D.}~\bibnamefont {Corsini}}, \bibinfo {author} {\bibfnamefont
  {M.}~\bibnamefont {Faverzani}}, \bibinfo {author} {\bibfnamefont
  {F.}~\bibnamefont {Gatti}}, \bibinfo {author} {\bibfnamefont
  {A.}~\bibnamefont {Giachero}}, \bibinfo {author} {\bibfnamefont
  {C.}~\bibnamefont {Gotti}}, \bibinfo {author} {\bibfnamefont
  {C.}~\bibnamefont {Kilbourne}}, \bibinfo {author} {\bibfnamefont
  {A.}~\bibnamefont {Kling}}, \bibinfo {author} {\bibfnamefont
  {M.}~\bibnamefont {Maino}}, \bibinfo {author} {\bibfnamefont
  {P.}~\bibnamefont {Manfrinetti}}, \bibinfo {author} {\bibfnamefont
  {A.}~\bibnamefont {Nucciotti}}, \bibinfo {author} {\bibfnamefont
  {G.}~\bibnamefont {Pessina}}, \bibinfo {author} {\bibfnamefont
  {G.}~\bibnamefont {Pizzigoni}}, \bibinfo {author} {\bibfnamefont {M.~R.}\
  \bibnamefont {Gomes}},\ and\ \bibinfo {author} {\bibfnamefont
  {M.}~\bibnamefont {Sisti}},\ }\bibfield  {title} {\bibinfo {title} {The
  status of the mare experiment with $^{187}${R}e and $^{163}${H}o isotopes},\
  }\href {https://doi.org/https://doi.org/10.1016/j.phpro.2014.12.037}
  {\bibfield  {journal} {\bibinfo  {journal} {Physics Procedia}\ }\textbf
  {\bibinfo {volume} {61}},\ \bibinfo {pages} {227} (\bibinfo {year} {2015})},\
  \bibinfo {note} {13th International Conference on Topics in Astroparticle and
  Underground Physics, TAUP 2013}\BibitemShut {NoStop}%
\bibitem [{\citenamefont {Dvornick\'y}\ \emph {et~al.}(2011)\citenamefont
  {Dvornick\'y}, \citenamefont {Muto}, \citenamefont {\ifmmode~\check{S}\else
  \v{S}\fi{}imkovic},\ and\ \citenamefont {Faessler}}]{RastislavPRC2011}%
  \BibitemOpen
  \bibfield  {author} {\bibinfo {author} {\bibfnamefont {R.}~\bibnamefont
  {Dvornick\'y}}, \bibinfo {author} {\bibfnamefont {K.}~\bibnamefont {Muto}},
  \bibinfo {author} {\bibfnamefont {F.}~\bibnamefont {\ifmmode~\check{S}\else
  \v{S}\fi{}imkovic}},\ and\ \bibinfo {author} {\bibfnamefont {A.}~\bibnamefont
  {Faessler}},\ }\bibfield  {title} {\bibinfo {title} {Absolute mass of
  neutrinos and the first unique forbidden $\ensuremath{\beta}$ decay of
  $^{187}\mathrm{Re}$},\ }\href {https://doi.org/10.1103/PhysRevC.83.045502}
  {\bibfield  {journal} {\bibinfo  {journal} {Phys. Rev. C}\ }\textbf {\bibinfo
  {volume} {83}},\ \bibinfo {pages} {045502} (\bibinfo {year}
  {2011})}\BibitemShut {NoStop}%
\bibitem [{\citenamefont {Myers}\ \emph {et~al.}(2015)\citenamefont {Myers},
  \citenamefont {Wagner}, \citenamefont {Kracke},\ and\ \citenamefont
  {Wesson}}]{MyersPRL2015}%
  \BibitemOpen
  \bibfield  {author} {\bibinfo {author} {\bibfnamefont {E.~G.}\ \bibnamefont
  {Myers}}, \bibinfo {author} {\bibfnamefont {A.}~\bibnamefont {Wagner}},
  \bibinfo {author} {\bibfnamefont {H.}~\bibnamefont {Kracke}},\ and\ \bibinfo
  {author} {\bibfnamefont {B.~A.}\ \bibnamefont {Wesson}},\ }\bibfield  {title}
  {\bibinfo {title} {Atomic masses of tritium and helium-3},\ }\href
  {https://doi.org/10.1103/PhysRevLett.114.013003} {\bibfield  {journal}
  {\bibinfo  {journal} {Phys. Rev. Lett.}\ }\textbf {\bibinfo {volume} {114}},\
  \bibinfo {pages} {013003} (\bibinfo {year} {2015})}\BibitemShut {NoStop}%
\bibitem [{\citenamefont {Filianin}\ \emph {et~al.}(2021)\citenamefont
  {Filianin}, \citenamefont {Lyu}, \citenamefont {Door}, \citenamefont {Blaum},
  \citenamefont {Huang}, \citenamefont {Haverkort}, \citenamefont {Indelicato},
  \citenamefont {Keitel}, \citenamefont {Kromer}, \citenamefont {Lange},
  \citenamefont {Novikov}, \citenamefont {Rischka}, \citenamefont
  {Sch\"ussler}, \citenamefont {Schweiger}, \citenamefont {Sturm},
  \citenamefont {Ulmer}, \citenamefont {Harman},\ and\ \citenamefont
  {Eliseev}}]{Qvalue187Re}%
  \BibitemOpen
  \bibfield  {author} {\bibinfo {author} {\bibfnamefont {P.}~\bibnamefont
  {Filianin}}, \bibinfo {author} {\bibfnamefont {C.}~\bibnamefont {Lyu}},
  \bibinfo {author} {\bibfnamefont {M.}~\bibnamefont {Door}}, \bibinfo {author}
  {\bibfnamefont {K.}~\bibnamefont {Blaum}}, \bibinfo {author} {\bibfnamefont
  {W.~J.}\ \bibnamefont {Huang}}, \bibinfo {author} {\bibfnamefont
  {M.}~\bibnamefont {Haverkort}}, \bibinfo {author} {\bibfnamefont
  {P.}~\bibnamefont {Indelicato}}, \bibinfo {author} {\bibfnamefont {C.~H.}\
  \bibnamefont {Keitel}}, \bibinfo {author} {\bibfnamefont {K.}~\bibnamefont
  {Kromer}}, \bibinfo {author} {\bibfnamefont {D.}~\bibnamefont {Lange}},
  \bibinfo {author} {\bibfnamefont {Y.~N.}\ \bibnamefont {Novikov}}, \bibinfo
  {author} {\bibfnamefont {A.}~\bibnamefont {Rischka}}, \bibinfo {author}
  {\bibfnamefont {R.~X.}\ \bibnamefont {Sch\"ussler}}, \bibinfo {author}
  {\bibfnamefont {C.}~\bibnamefont {Schweiger}}, \bibinfo {author}
  {\bibfnamefont {S.}~\bibnamefont {Sturm}}, \bibinfo {author} {\bibfnamefont
  {S.}~\bibnamefont {Ulmer}}, \bibinfo {author} {\bibfnamefont
  {Z.}~\bibnamefont {Harman}},\ and\ \bibinfo {author} {\bibfnamefont
  {S.}~\bibnamefont {Eliseev}},\ }\bibfield  {title} {\bibinfo {title} {Direct
  $q$-value determination of the ${\ensuremath{\beta}}^{\ensuremath{-}}$ decay
  of $^{187}\mathrm{Re}$},\ }\href
  {https://doi.org/10.1103/PhysRevLett.127.072502} {\bibfield  {journal}
  {\bibinfo  {journal} {Phys. Rev. Lett.}\ }\textbf {\bibinfo {volume} {127}},\
  \bibinfo {pages} {072502} (\bibinfo {year} {2021})}\BibitemShut {NoStop}%
\bibitem [{\citenamefont {Faverzani}\ \emph {et~al.}(2016)\citenamefont
  {Faverzani} \emph {et~al.}}]{HOLMESexperimentJLTP2016}%
  \BibitemOpen
  \bibfield  {author} {\bibinfo {author} {\bibfnamefont {M.}~\bibnamefont
  {Faverzani}} \emph {et~al.},\ }\bibfield  {title} {\bibinfo {title} {The
  {HOLMES} experiment},\ }\bibfield  {journal} {\bibinfo  {journal} {Journal of
  Low Temperature Physics}\ }\textbf {\bibinfo {volume} {184}},\ \href
  {https://doi.org/10.1007/s10909-016-1540-x} {10.1007/s10909-016-1540-x}
  (\bibinfo {year} {2016})\BibitemShut {NoStop}%
\bibitem [{\citenamefont {Croce}\ \emph {et~al.}(2016)\citenamefont {Croce}
  \emph {et~al.}}]{NuMECsexperimentJLTP2016}%
  \BibitemOpen
  \bibfield  {author} {\bibinfo {author} {\bibfnamefont {M.~P.}\ \bibnamefont
  {Croce}} \emph {et~al.},\ }\bibfield  {title} {\bibinfo {title} {Development
  of holmium-163 electron-capture spectroscopy with transition-edge sensors},\
  }\bibfield  {journal} {\bibinfo  {journal} {Journal of Low Temperature
  Physics}\ }\textbf {\bibinfo {volume} {184}},\ \href
  {https://doi.org/10.1007/s10909-015-1451-2} {10.1007/s10909-015-1451-2}
  (\bibinfo {year} {2016})\BibitemShut {NoStop}%
\bibitem [{\citenamefont {Gastaldo}\ \emph {et~al.}(2017)\citenamefont
  {Gastaldo} \emph {et~al.}}]{ECHoTEPJST2017}%
  \BibitemOpen
  \bibfield  {author} {\bibinfo {author} {\bibfnamefont {L.}~\bibnamefont
  {Gastaldo}} \emph {et~al.},\ }\bibfield  {title} {\bibinfo {title} {The
  electron capture in $^{163}${H}o experiment – {ECH}o},\ }\bibfield
  {journal} {\bibinfo  {journal} {The European Physical Journal Special
  Topics}\ }\textbf {\bibinfo {volume} {226}},\ \href
  {https://doi.org/10.1140/epjst/e2017-70071-y} {10.1140/epjst/e2017-70071-y}
  (\bibinfo {year} {2017})\BibitemShut {NoStop}%
\bibitem [{\citenamefont {Eliseev}\ \emph {et~al.}(2015)\citenamefont
  {Eliseev}, \citenamefont {Blaum}, \citenamefont {Block}, \citenamefont
  {Chenmarev}, \citenamefont {Dorrer}, \citenamefont {D\"ullmann},
  \citenamefont {Enss}, \citenamefont {Filianin}, \citenamefont {Gastaldo},
  \citenamefont {Goncharov}, \citenamefont {K\"oster}, \citenamefont
  {Lautenschl\"ager}, \citenamefont {Novikov}, \citenamefont {Rischka},
  \citenamefont {Sch\"ussler}, \citenamefont {Schweikhard},\ and\ \citenamefont
  {T\"urler}}]{EliseevPRL2016}%
  \BibitemOpen
  \bibfield  {author} {\bibinfo {author} {\bibfnamefont {S.}~\bibnamefont
  {Eliseev}}, \bibinfo {author} {\bibfnamefont {K.}~\bibnamefont {Blaum}},
  \bibinfo {author} {\bibfnamefont {M.}~\bibnamefont {Block}}, \bibinfo
  {author} {\bibfnamefont {S.}~\bibnamefont {Chenmarev}}, \bibinfo {author}
  {\bibfnamefont {H.}~\bibnamefont {Dorrer}}, \bibinfo {author} {\bibfnamefont
  {C.~E.}\ \bibnamefont {D\"ullmann}}, \bibinfo {author} {\bibfnamefont
  {C.}~\bibnamefont {Enss}}, \bibinfo {author} {\bibfnamefont {P.~E.}\
  \bibnamefont {Filianin}}, \bibinfo {author} {\bibfnamefont {L.}~\bibnamefont
  {Gastaldo}}, \bibinfo {author} {\bibfnamefont {M.}~\bibnamefont {Goncharov}},
  \bibinfo {author} {\bibfnamefont {U.}~\bibnamefont {K\"oster}}, \bibinfo
  {author} {\bibfnamefont {F.}~\bibnamefont {Lautenschl\"ager}}, \bibinfo
  {author} {\bibfnamefont {Y.~N.}\ \bibnamefont {Novikov}}, \bibinfo {author}
  {\bibfnamefont {A.}~\bibnamefont {Rischka}}, \bibinfo {author} {\bibfnamefont
  {R.~X.}\ \bibnamefont {Sch\"ussler}}, \bibinfo {author} {\bibfnamefont
  {L.}~\bibnamefont {Schweikhard}},\ and\ \bibinfo {author} {\bibfnamefont
  {A.}~\bibnamefont {T\"urler}},\ }\bibfield  {title} {\bibinfo {title} {Direct
  measurement of the mass difference of $^{163}\mathrm{Ho}$ and
  $^{163}\mathrm{Dy}$ solves the $q$-value puzzle for the neutrino mass
  determination},\ }\href {https://doi.org/10.1103/PhysRevLett.115.062501}
  {\bibfield  {journal} {\bibinfo  {journal} {Phys. Rev. Lett.}\ }\textbf
  {\bibinfo {volume} {115}},\ \bibinfo {pages} {062501} (\bibinfo {year}
  {2015})}\BibitemShut {NoStop}%
\bibitem [{\citenamefont {Gamage}\ \emph {et~al.}(2019)\citenamefont {Gamage},
  \citenamefont {Bhandari}, \citenamefont {Horana~Gamage}, \citenamefont
  {Sandler},\ and\ \citenamefont {Redshaw}}]{GamageHY2019}%
  \BibitemOpen
  \bibfield  {author} {\bibinfo {author} {\bibfnamefont {N.~D.}\ \bibnamefont
  {Gamage}}, \bibinfo {author} {\bibfnamefont {R.}~\bibnamefont {Bhandari}},
  \bibinfo {author} {\bibfnamefont {M.}~\bibnamefont {Horana~Gamage}}, \bibinfo
  {author} {\bibfnamefont {R.}~\bibnamefont {Sandler}},\ and\ \bibinfo {author}
  {\bibfnamefont {M.}~\bibnamefont {Redshaw}},\ }\bibfield  {title} {\bibinfo
  {title} {Identification and investigation of possible ultra-low {Q} value
  $\beta$ decay candidates},\ }\bibfield  {journal} {\bibinfo  {journal}
  {Hyperfine Interactions}\ }\textbf {\bibinfo {volume} {240}},\ \href
  {https://doi.org/10.1007/s10751-019-1588-5} {10.1007/s10751-019-1588-5}
  (\bibinfo {year} {2019})\BibitemShut {NoStop}%
\bibitem [{\citenamefont {Redshaw}(2023)}]{RedshawEPJA2023}%
  \BibitemOpen
  \bibfield  {author} {\bibinfo {author} {\bibfnamefont {M.}~\bibnamefont
  {Redshaw}},\ }\bibfield  {title} {\bibinfo {title} {Precise q value
  determinations for forbidden and low energy {$\beta$}-decays using penning
  trap mass spectrometry},\ }\href
  {https://doi.org/10.1140/epja/s10050-023-00925-9} {\bibfield  {journal}
  {\bibinfo  {journal} {The European Physical Journal A}\ }\textbf {\bibinfo
  {volume} {59}},\ \bibinfo {pages} {18} (\bibinfo {year} {2023})}\BibitemShut
  {NoStop}%
\bibitem [{\citenamefont {Keblbeck}\ \emph {et~al.}(2023)\citenamefont
  {Keblbeck}, \citenamefont {Bhandari}, \citenamefont {Gamage}, \citenamefont
  {Horana~Gamage}, \citenamefont {Leach}, \citenamefont {Mougeot},\ and\
  \citenamefont {Redshaw}}]{KeblbeckPRC2023}%
  \BibitemOpen
  \bibfield  {author} {\bibinfo {author} {\bibfnamefont {D.~K.}\ \bibnamefont
  {Keblbeck}}, \bibinfo {author} {\bibfnamefont {R.}~\bibnamefont {Bhandari}},
  \bibinfo {author} {\bibfnamefont {N.~D.}\ \bibnamefont {Gamage}}, \bibinfo
  {author} {\bibfnamefont {M.}~\bibnamefont {Horana~Gamage}}, \bibinfo {author}
  {\bibfnamefont {K.~G.}\ \bibnamefont {Leach}}, \bibinfo {author}
  {\bibfnamefont {X.}~\bibnamefont {Mougeot}},\ and\ \bibinfo {author}
  {\bibfnamefont {M.}~\bibnamefont {Redshaw}},\ }\bibfield  {title} {\bibinfo
  {title} {Updated evaluation of potential ultralow $q$-value
  $\ensuremath{\beta}$-decay candidates},\ }\href
  {https://doi.org/10.1103/PhysRevC.107.015504} {\bibfield  {journal} {\bibinfo
   {journal} {Phys. Rev. C}\ }\textbf {\bibinfo {volume} {107}},\ \bibinfo
  {pages} {015504} (\bibinfo {year} {2023})}\BibitemShut {NoStop}%
\bibitem [{\citenamefont {Aker}\ \emph {et~al.}(2022)\citenamefont {Aker},
  \citenamefont {Beglarian}, \citenamefont {Behrens}, \citenamefont {Berlev},
  \citenamefont {Besserer}, \citenamefont {Bieringer} \emph
  {et~al.}}]{KATRIN-N2022}%
  \BibitemOpen
  \bibfield  {author} {\bibinfo {author} {\bibfnamefont {M.}~\bibnamefont
  {Aker}}, \bibinfo {author} {\bibfnamefont {A.}~\bibnamefont {Beglarian}},
  \bibinfo {author} {\bibfnamefont {J.}~\bibnamefont {Behrens}}, \bibinfo
  {author} {\bibfnamefont {A.}~\bibnamefont {Berlev}}, \bibinfo {author}
  {\bibfnamefont {U.}~\bibnamefont {Besserer}}, \bibinfo {author}
  {\bibfnamefont {F.}~\bibnamefont {Bieringer}, \bibfnamefont {B.~Block}},
  \emph {et~al.} (\bibinfo {collaboration} {The KATRIN Collaboration}),\
  }\bibfield  {title} {\bibinfo {title} {Direct neutrino-mass measurement with
  sub-electronvolt sensitivity},\ }\href
  {https://doi.org/10.1038/s41567-021-01463-1} {\bibfield  {journal} {\bibinfo
  {journal} {Nature Physics}\ }\textbf {\bibinfo {volume} {18}},\ \bibinfo
  {pages} {160} (\bibinfo {year} {2022})}\BibitemShut {NoStop}%
\bibitem [{\citenamefont {Aseev}\ \emph {et~al.}(2011)\citenamefont {Aseev},
  \citenamefont {Belesev}, \citenamefont {Berlev}, \citenamefont {Geraskin},
  \citenamefont {Golubev}, \citenamefont {Likhovid}, \citenamefont {Lobashev},
  \citenamefont {Nozik}, \citenamefont {Pantuev}, \citenamefont {Parfenov},
  \citenamefont {Skasyrskaya}, \citenamefont {Tkachov},\ and\ \citenamefont
  {Zadorozhny}}]{TroitskPRD2011}%
  \BibitemOpen
  \bibfield  {author} {\bibinfo {author} {\bibfnamefont {V.~N.}\ \bibnamefont
  {Aseev}}, \bibinfo {author} {\bibfnamefont {A.~I.}\ \bibnamefont {Belesev}},
  \bibinfo {author} {\bibfnamefont {A.~I.}\ \bibnamefont {Berlev}}, \bibinfo
  {author} {\bibfnamefont {E.~V.}\ \bibnamefont {Geraskin}}, \bibinfo {author}
  {\bibfnamefont {A.~A.}\ \bibnamefont {Golubev}}, \bibinfo {author}
  {\bibfnamefont {N.~A.}\ \bibnamefont {Likhovid}}, \bibinfo {author}
  {\bibfnamefont {V.~M.}\ \bibnamefont {Lobashev}}, \bibinfo {author}
  {\bibfnamefont {A.~A.}\ \bibnamefont {Nozik}}, \bibinfo {author}
  {\bibfnamefont {V.~S.}\ \bibnamefont {Pantuev}}, \bibinfo {author}
  {\bibfnamefont {V.~I.}\ \bibnamefont {Parfenov}}, \bibinfo {author}
  {\bibfnamefont {A.~K.}\ \bibnamefont {Skasyrskaya}}, \bibinfo {author}
  {\bibfnamefont {F.~V.}\ \bibnamefont {Tkachov}},\ and\ \bibinfo {author}
  {\bibfnamefont {S.~V.}\ \bibnamefont {Zadorozhny}},\ }\bibfield  {title}
  {\bibinfo {title} {Upper limit on the electron antineutrino mass from the
  troitsk experiment},\ }\href {https://doi.org/10.1103/PhysRevD.84.112003}
  {\bibfield  {journal} {\bibinfo  {journal} {Phys. Rev. D}\ }\textbf {\bibinfo
  {volume} {84}},\ \bibinfo {pages} {112003} (\bibinfo {year}
  {2011})}\BibitemShut {NoStop}%
\bibitem [{\citenamefont {Kraus}\ \emph {et~al.}(2005)\citenamefont {Kraus}
  \emph {et~al.}}]{MainzEPJC2005}%
  \BibitemOpen
  \bibfield  {author} {\bibinfo {author} {\bibfnamefont {C.~P.}\ \bibnamefont
  {Kraus}} \emph {et~al.},\ }\bibfield  {title} {\bibinfo {title} {Final
  results from phase ii of the mainz neutrino mass searchin tritium
  ${\beta}$decay},\ }\bibfield  {journal} {\bibinfo  {journal} {The European
  Physical Journal C - Particles and Fields}\ }\textbf {\bibinfo {volume}
  {40}},\ \href {https://doi.org/10.1140/epjc/s2005-02139-7}
  {10.1140/epjc/s2005-02139-7} (\bibinfo {year} {2005})\BibitemShut {NoStop}%
\bibitem [{\citenamefont {Ashtari~Esfahani}\ \emph {et~al.}(2023)\citenamefont
  {Ashtari~Esfahani}, \citenamefont {B\"oser}, \citenamefont {Buzinsky},
  \citenamefont {Carmona-Benitez}, \citenamefont {Claessens}, \citenamefont
  {de~Viveiros}, \citenamefont {Doe}, \citenamefont {Fertl}, \citenamefont
  {Formaggio}, \citenamefont {Gaison}, \citenamefont {Gladstone}, \citenamefont
  {Grando}, \citenamefont {Guigue}, \citenamefont {Hartse}, \citenamefont
  {Heeger}, \citenamefont {Huyan}, \citenamefont {Johnston}, \citenamefont
  {Jones}, \citenamefont {Kazkaz}, \citenamefont {LaRoque}, \citenamefont {Li},
  \citenamefont {Lindman}, \citenamefont {Machado}, \citenamefont {Marsteller},
  \citenamefont {Matth\'e}, \citenamefont {Mohiuddin}, \citenamefont {Monreal},
  \citenamefont {Mueller}, \citenamefont {Nikkel}, \citenamefont {Novitski},
  \citenamefont {Oblath}, \citenamefont {Pe\~na}, \citenamefont {Pettus},
  \citenamefont {Reimann}, \citenamefont {Robertson}, \citenamefont {Rosa
  De~Jes\'us}, \citenamefont {Rybka}, \citenamefont {Salda\~na}, \citenamefont
  {Schram}, \citenamefont {Slocum}, \citenamefont {Stachurska}, \citenamefont
  {Sun}, \citenamefont {Surukuchi}, \citenamefont {Tedeschi}, \citenamefont
  {Telles}, \citenamefont {Thomas}, \citenamefont {Thomas}, \citenamefont
  {Thorne}, \citenamefont {Th\"ummler}, \citenamefont {Tvrznikova},
  \citenamefont {Van De~Pontseele}, \citenamefont {VanDevender}, \citenamefont
  {Weintroub}, \citenamefont {Weiss}, \citenamefont {Wendler}, \citenamefont
  {Young}, \citenamefont {Zayas},\ and\ \citenamefont
  {Ziegler}}]{Project8PRL2023}%
  \BibitemOpen
  \bibfield  {author} {\bibinfo {author} {\bibfnamefont {A.}~\bibnamefont
  {Ashtari~Esfahani}}, \bibinfo {author} {\bibfnamefont {S.}~\bibnamefont
  {B\"oser}}, \bibinfo {author} {\bibfnamefont {N.}~\bibnamefont {Buzinsky}},
  \bibinfo {author} {\bibfnamefont {M.~C.}\ \bibnamefont {Carmona-Benitez}},
  \bibinfo {author} {\bibfnamefont {C.}~\bibnamefont {Claessens}}, \bibinfo
  {author} {\bibfnamefont {L.}~\bibnamefont {de~Viveiros}}, \bibinfo {author}
  {\bibfnamefont {P.~J.}\ \bibnamefont {Doe}}, \bibinfo {author} {\bibfnamefont
  {M.}~\bibnamefont {Fertl}}, \bibinfo {author} {\bibfnamefont {J.~A.}\
  \bibnamefont {Formaggio}}, \bibinfo {author} {\bibfnamefont {J.~K.}\
  \bibnamefont {Gaison}}, \bibinfo {author} {\bibfnamefont {L.}~\bibnamefont
  {Gladstone}}, \bibinfo {author} {\bibfnamefont {M.}~\bibnamefont {Grando}},
  \bibinfo {author} {\bibfnamefont {M.}~\bibnamefont {Guigue}}, \bibinfo
  {author} {\bibfnamefont {J.}~\bibnamefont {Hartse}}, \bibinfo {author}
  {\bibfnamefont {K.~M.}\ \bibnamefont {Heeger}}, \bibinfo {author}
  {\bibfnamefont {X.}~\bibnamefont {Huyan}}, \bibinfo {author} {\bibfnamefont
  {J.}~\bibnamefont {Johnston}}, \bibinfo {author} {\bibfnamefont {A.~M.}\
  \bibnamefont {Jones}}, \bibinfo {author} {\bibfnamefont {K.}~\bibnamefont
  {Kazkaz}}, \bibinfo {author} {\bibfnamefont {B.~H.}\ \bibnamefont {LaRoque}},
  \bibinfo {author} {\bibfnamefont {M.}~\bibnamefont {Li}}, \bibinfo {author}
  {\bibfnamefont {A.}~\bibnamefont {Lindman}}, \bibinfo {author} {\bibfnamefont
  {E.}~\bibnamefont {Machado}}, \bibinfo {author} {\bibfnamefont
  {A.}~\bibnamefont {Marsteller}}, \bibinfo {author} {\bibfnamefont
  {C.}~\bibnamefont {Matth\'e}}, \bibinfo {author} {\bibfnamefont
  {R.}~\bibnamefont {Mohiuddin}}, \bibinfo {author} {\bibfnamefont
  {B.}~\bibnamefont {Monreal}}, \bibinfo {author} {\bibfnamefont
  {R.}~\bibnamefont {Mueller}}, \bibinfo {author} {\bibfnamefont {J.~A.}\
  \bibnamefont {Nikkel}}, \bibinfo {author} {\bibfnamefont {E.}~\bibnamefont
  {Novitski}}, \bibinfo {author} {\bibfnamefont {N.~S.}\ \bibnamefont
  {Oblath}}, \bibinfo {author} {\bibfnamefont {J.~I.}\ \bibnamefont {Pe\~na}},
  \bibinfo {author} {\bibfnamefont {W.}~\bibnamefont {Pettus}}, \bibinfo
  {author} {\bibfnamefont {R.}~\bibnamefont {Reimann}}, \bibinfo {author}
  {\bibfnamefont {R.~G.~H.}\ \bibnamefont {Robertson}}, \bibinfo {author}
  {\bibfnamefont {D.}~\bibnamefont {Rosa De~Jes\'us}}, \bibinfo {author}
  {\bibfnamefont {G.}~\bibnamefont {Rybka}}, \bibinfo {author} {\bibfnamefont
  {L.}~\bibnamefont {Salda\~na}}, \bibinfo {author} {\bibfnamefont
  {M.}~\bibnamefont {Schram}}, \bibinfo {author} {\bibfnamefont {P.~L.}\
  \bibnamefont {Slocum}}, \bibinfo {author} {\bibfnamefont {J.}~\bibnamefont
  {Stachurska}}, \bibinfo {author} {\bibfnamefont {Y.-H.}\ \bibnamefont {Sun}},
  \bibinfo {author} {\bibfnamefont {P.~T.}\ \bibnamefont {Surukuchi}}, \bibinfo
  {author} {\bibfnamefont {J.~R.}\ \bibnamefont {Tedeschi}}, \bibinfo {author}
  {\bibfnamefont {A.~B.}\ \bibnamefont {Telles}}, \bibinfo {author}
  {\bibfnamefont {F.}~\bibnamefont {Thomas}}, \bibinfo {author} {\bibfnamefont
  {M.}~\bibnamefont {Thomas}}, \bibinfo {author} {\bibfnamefont {L.~A.}\
  \bibnamefont {Thorne}}, \bibinfo {author} {\bibfnamefont {T.}~\bibnamefont
  {Th\"ummler}}, \bibinfo {author} {\bibfnamefont {L.}~\bibnamefont
  {Tvrznikova}}, \bibinfo {author} {\bibfnamefont {W.}~\bibnamefont {Van
  De~Pontseele}}, \bibinfo {author} {\bibfnamefont {B.~A.}\ \bibnamefont
  {VanDevender}}, \bibinfo {author} {\bibfnamefont {J.}~\bibnamefont
  {Weintroub}}, \bibinfo {author} {\bibfnamefont {T.~E.}\ \bibnamefont
  {Weiss}}, \bibinfo {author} {\bibfnamefont {T.}~\bibnamefont {Wendler}},
  \bibinfo {author} {\bibfnamefont {A.}~\bibnamefont {Young}}, \bibinfo
  {author} {\bibfnamefont {E.}~\bibnamefont {Zayas}},\ and\ \bibinfo {author}
  {\bibfnamefont {A.}~\bibnamefont {Ziegler}} (\bibinfo {collaboration}
  {Project 8 Collaboration}),\ }\bibfield  {title} {\bibinfo {title} {Tritium
  beta spectrum measurement and neutrino mass limit from cyclotron radiation
  emission spectroscopy},\ }\href
  {https://doi.org/10.1103/PhysRevLett.131.102502} {\bibfield  {journal}
  {\bibinfo  {journal} {Phys. Rev. Lett.}\ }\textbf {\bibinfo {volume} {131}},\
  \bibinfo {pages} {102502} (\bibinfo {year} {2023})}\BibitemShut {NoStop}%
\bibitem [{\citenamefont {Gatti}\ \emph {et~al.}(1999)\citenamefont {Gatti},
  \citenamefont {Fontanelli}, \citenamefont {Galeazzi}, \citenamefont {Swift},\
  and\ \citenamefont {Vitale}}]{GattiN1999}%
  \BibitemOpen
  \bibfield  {author} {\bibinfo {author} {\bibfnamefont {F.}~\bibnamefont
  {Gatti}}, \bibinfo {author} {\bibfnamefont {F.}~\bibnamefont {Fontanelli}},
  \bibinfo {author} {\bibfnamefont {M.}~\bibnamefont {Galeazzi}}, \bibinfo
  {author} {\bibfnamefont {A.~M.}\ \bibnamefont {Swift}},\ and\ \bibinfo
  {author} {\bibfnamefont {S.}~\bibnamefont {Vitale}},\ }\bibfield  {title}
  {\bibinfo {title} {Detection of environmental fine structure in the
  low-energy $\beta$-decay spectrum of $^{187}$re},\ }\href
  {https://doi.org/10.1038/16414} {\bibfield  {journal} {\bibinfo  {journal}
  {Nature}\ }\textbf {\bibinfo {volume} {397}},\ \bibinfo {pages} {137}
  (\bibinfo {year} {1999})}\BibitemShut {NoStop}%
\bibitem [{\citenamefont {Arnaboldi}\ \emph {et~al.}(2006)\citenamefont
  {Arnaboldi}, \citenamefont {Benedek}, \citenamefont {Brofferio},
  \citenamefont {Capelli}, \citenamefont {Capozzi}, \citenamefont {Cremonesi},
  \citenamefont {Filipponi}, \citenamefont {Fiorini}, \citenamefont {Giuliani},
  \citenamefont {Monfardini}, \citenamefont {Nucciotti}, \citenamefont {Pavan},
  \citenamefont {Pedretti}, \citenamefont {Pessina}, \citenamefont {Pirro},
  \citenamefont {Previtali},\ and\ \citenamefont {Sisti}}]{MARE-PRL2006}%
  \BibitemOpen
  \bibfield  {author} {\bibinfo {author} {\bibfnamefont {C.}~\bibnamefont
  {Arnaboldi}}, \bibinfo {author} {\bibfnamefont {G.}~\bibnamefont {Benedek}},
  \bibinfo {author} {\bibfnamefont {C.}~\bibnamefont {Brofferio}}, \bibinfo
  {author} {\bibfnamefont {S.}~\bibnamefont {Capelli}}, \bibinfo {author}
  {\bibfnamefont {F.}~\bibnamefont {Capozzi}}, \bibinfo {author} {\bibfnamefont
  {O.}~\bibnamefont {Cremonesi}}, \bibinfo {author} {\bibfnamefont
  {A.}~\bibnamefont {Filipponi}}, \bibinfo {author} {\bibfnamefont
  {E.}~\bibnamefont {Fiorini}}, \bibinfo {author} {\bibfnamefont
  {A.}~\bibnamefont {Giuliani}}, \bibinfo {author} {\bibfnamefont
  {A.}~\bibnamefont {Monfardini}}, \bibinfo {author} {\bibfnamefont
  {A.}~\bibnamefont {Nucciotti}}, \bibinfo {author} {\bibfnamefont
  {M.}~\bibnamefont {Pavan}}, \bibinfo {author} {\bibfnamefont
  {M.}~\bibnamefont {Pedretti}}, \bibinfo {author} {\bibfnamefont
  {G.}~\bibnamefont {Pessina}}, \bibinfo {author} {\bibfnamefont
  {S.}~\bibnamefont {Pirro}}, \bibinfo {author} {\bibfnamefont
  {E.}~\bibnamefont {Previtali}},\ and\ \bibinfo {author} {\bibfnamefont
  {M.}~\bibnamefont {Sisti}},\ }\bibfield  {title} {\bibinfo {title}
  {Measurement of the $p$ to $s$ wave branching ratio of $^{187}\mathrm{Re}$
  $\ensuremath{\beta}$ decay from beta environmental fine structure},\ }\href
  {https://doi.org/10.1103/PhysRevLett.96.042503} {\bibfield  {journal}
  {\bibinfo  {journal} {Phys. Rev. Lett.}\ }\textbf {\bibinfo {volume} {96}},\
  \bibinfo {pages} {042503} (\bibinfo {year} {2006})}\BibitemShut {NoStop}%
\bibitem [{\citenamefont {Nucciotti}(2008)}]{NucciottiJLTP2008}%
  \BibitemOpen
  \bibfield  {author} {\bibinfo {author} {\bibfnamefont {A.}~\bibnamefont
  {Nucciotti}},\ }\bibfield  {title} {\bibinfo {title} {The {MARE} project},\
  }\href {https://doi.org/10.1007/s10909-008-9718-5} {\bibfield  {journal}
  {\bibinfo  {journal} {Journal of Low Temperature Physics}\ }\textbf {\bibinfo
  {volume} {151}},\ \bibinfo {pages} {597} (\bibinfo {year}
  {2008})}\BibitemShut {NoStop}%
\bibitem [{\citenamefont {Formaggio}\ \emph {et~al.}(2021)\citenamefont
  {Formaggio}, \citenamefont {{de Gouvêa}},\ and\ \citenamefont
  {Robertson}}]{FormaggioPR2021}%
  \BibitemOpen
  \bibfield  {author} {\bibinfo {author} {\bibfnamefont {J.~A.}\ \bibnamefont
  {Formaggio}}, \bibinfo {author} {\bibfnamefont {A.~L.~C.}\ \bibnamefont {{de
  Gouvêa}}},\ and\ \bibinfo {author} {\bibfnamefont {R.~H.}\ \bibnamefont
  {Robertson}},\ }\bibfield  {title} {\bibinfo {title} {Direct measurements of
  neutrino mass},\ }\href
  {https://doi.org/https://doi.org/10.1016/j.physrep.2021.02.002} {\bibfield
  {journal} {\bibinfo  {journal} {Physics Reports}\ }\textbf {\bibinfo {volume}
  {914}},\ \bibinfo {pages} {1} (\bibinfo {year} {2021})},\ \bibinfo {note}
  {direct measurements of neutrino mass}\BibitemShut {NoStop}%
\bibitem [{\citenamefont {Wilkinson}(1991)}]{WilkinsonNPA1991}%
  \BibitemOpen
  \bibfield  {author} {\bibinfo {author} {\bibfnamefont {D.}~\bibnamefont
  {Wilkinson}},\ }\bibfield  {title} {\bibinfo {title} {Small terms in the
  beta-decay spectrum of tritium},\ }\href
  {https://doi.org/https://doi.org/10.1016/0375-9474(91)90301-L} {\bibfield
  {journal} {\bibinfo  {journal} {Nuclear Physics A}\ }\textbf {\bibinfo
  {volume} {526}},\ \bibinfo {pages} {131} (\bibinfo {year}
  {1991})}\BibitemShut {NoStop}%
\bibitem [{\citenamefont {Kleesiek}\ \emph {et~al.}(2019)\citenamefont
  {Kleesiek}, \citenamefont {Behrens}, \citenamefont {Drexlin}, \citenamefont
  {Eitel}, \citenamefont {Erhard}, \citenamefont {Formaggio}, \citenamefont
  {Gl{\"u}ck}, \citenamefont {Groh}, \citenamefont {H{\"o}tzel}, \citenamefont
  {Mertens}, \citenamefont {Poon}, \citenamefont {Weinheimer},\ and\
  \citenamefont {Valerius}}]{Kleesiek2019}%
  \BibitemOpen
  \bibfield  {author} {\bibinfo {author} {\bibfnamefont {M.}~\bibnamefont
  {Kleesiek}}, \bibinfo {author} {\bibfnamefont {J.}~\bibnamefont {Behrens}},
  \bibinfo {author} {\bibfnamefont {G.}~\bibnamefont {Drexlin}}, \bibinfo
  {author} {\bibfnamefont {K.}~\bibnamefont {Eitel}}, \bibinfo {author}
  {\bibfnamefont {M.}~\bibnamefont {Erhard}}, \bibinfo {author} {\bibfnamefont
  {J.~A.}\ \bibnamefont {Formaggio}}, \bibinfo {author} {\bibfnamefont
  {F.}~\bibnamefont {Gl{\"u}ck}}, \bibinfo {author} {\bibfnamefont
  {S.}~\bibnamefont {Groh}}, \bibinfo {author} {\bibfnamefont {M.}~\bibnamefont
  {H{\"o}tzel}}, \bibinfo {author} {\bibfnamefont {S.}~\bibnamefont {Mertens}},
  \bibinfo {author} {\bibfnamefont {A.~W.~P.}\ \bibnamefont {Poon}}, \bibinfo
  {author} {\bibfnamefont {C.}~\bibnamefont {Weinheimer}},\ and\ \bibinfo
  {author} {\bibfnamefont {K.}~\bibnamefont {Valerius}},\ }\bibfield  {title}
  {\bibinfo {title} {{$\beta$}-decay spectrum, response function and
  statistical model for neutrino mass measurements with the katrin
  experiment},\ }\href {https://doi.org/10.1140/epjc/s10052-019-6686-7}
  {\bibfield  {journal} {\bibinfo  {journal} {The European Physical Journal C}\
  }\textbf {\bibinfo {volume} {79}},\ \bibinfo {pages} {204} (\bibinfo {year}
  {2019})}\BibitemShut {NoStop}%
\bibitem [{\citenamefont {Rose}(1961)}]{RoseBook1961}%
  \BibitemOpen
  \bibfield  {author} {\bibinfo {author} {\bibfnamefont {M.~E.}\ \bibnamefont
  {Rose}},\ }\href@noop {} {\emph {\bibinfo {title} {Relativistic {E}lectron
  {T}heory}}}\ (\bibinfo  {publisher} {John Wiley and Sons},\ \bibinfo {year}
  {1961})\BibitemShut {NoStop}%
\bibitem [{\citenamefont {Salvat}\ and\ \citenamefont
  {Fernández-Varea}(2019)}]{SalvatCPC2019}%
  \BibitemOpen
  \bibfield  {author} {\bibinfo {author} {\bibfnamefont {F.}~\bibnamefont
  {Salvat}}\ and\ \bibinfo {author} {\bibfnamefont {J.~M.}\ \bibnamefont
  {Fernández-Varea}},\ }\bibfield  {title} {\bibinfo {title} {{RADIAL}: A
  fortran subroutine package for the solution of the radial {S}chr\"odinger and
  {D}irac wave equations},\ }\href
  {https://doi.org/https://doi.org/10.1016/j.cpc.2019.02.011} {\bibfield
  {journal} {\bibinfo  {journal} {Computer Physics Communications}\ }\textbf
  {\bibinfo {volume} {240}},\ \bibinfo {pages} {165} (\bibinfo {year}
  {2019})}\BibitemShut {NoStop}%
\bibitem [{\citenamefont {Doi}\ \emph {et~al.}(1985)\citenamefont {Doi},
  \citenamefont {Kotani},\ and\ \citenamefont {Takasugi}}]{DoiPTPS1985}%
  \BibitemOpen
  \bibfield  {author} {\bibinfo {author} {\bibfnamefont {M.}~\bibnamefont
  {Doi}}, \bibinfo {author} {\bibfnamefont {T.}~\bibnamefont {Kotani}},\ and\
  \bibinfo {author} {\bibfnamefont {E.}~\bibnamefont {Takasugi}},\ }\bibfield
  {title} {\bibinfo {title} {{Double Beta Decay and Majorana Neutrino}},\
  }\href {https://doi.org/10.1143/PTPS.83.1} {\bibfield  {journal} {\bibinfo
  {journal} {Progress of Theoretical Physics Supplement}\ }\textbf {\bibinfo
  {volume} {83}},\ \bibinfo {pages} {1} (\bibinfo {year} {1985})},\ \Eprint
  {https://arxiv.org/abs/https://academic.oup.com/ptps/article-pdf/doi/10.1143/PTPS.83.1/5227078/83-1.pdf}
  {https://academic.oup.com/ptps/article-pdf/doi/10.1143/PTPS.83.1/5227078/83-1.pdf}
  \BibitemShut {NoStop}%
\bibitem [{\citenamefont {Beresteckij}\ \emph {et~al.}(1989)\citenamefont
  {Beresteckij}, \citenamefont {Lifshitz},\ and\ \citenamefont
  {Pitaevskij}}]{BeresteckijBook1989}%
  \BibitemOpen
  \bibfield  {author} {\bibinfo {author} {\bibfnamefont {V.~B.}\ \bibnamefont
  {Beresteckij}}, \bibinfo {author} {\bibfnamefont {E.~M.}\ \bibnamefont
  {Lifshitz}},\ and\ \bibinfo {author} {\bibfnamefont {L.~P.}\ \bibnamefont
  {Pitaevskij}},\ }\href@noop {} {\emph {\bibinfo {title} {Quantum
  {E}lectrodynamics}}}\ (\bibinfo  {publisher} {Nauka},\ \bibinfo {year}
  {1989})\BibitemShut {NoStop}%
\bibitem [{\citenamefont {Hahn}\ \emph {et~al.}(1956)\citenamefont {Hahn},
  \citenamefont {Ravenhall},\ and\ \citenamefont {Hofstadter}}]{HahnPR1956}%
  \BibitemOpen
  \bibfield  {author} {\bibinfo {author} {\bibfnamefont {B.}~\bibnamefont
  {Hahn}}, \bibinfo {author} {\bibfnamefont {D.~G.}\ \bibnamefont
  {Ravenhall}},\ and\ \bibinfo {author} {\bibfnamefont {R.}~\bibnamefont
  {Hofstadter}},\ }\bibfield  {title} {\bibinfo {title} {High-energy electron
  scattering and the charge distributions of selected nuclei},\ }\href
  {https://doi.org/10.1103/PhysRev.101.1131} {\bibfield  {journal} {\bibinfo
  {journal} {Phys. Rev.}\ }\textbf {\bibinfo {volume} {101}},\ \bibinfo {pages}
  {1131} (\bibinfo {year} {1956})}\BibitemShut {NoStop}%
\bibitem [{\citenamefont {Slater}(1951)}]{SlaterPR1951}%
  \BibitemOpen
  \bibfield  {author} {\bibinfo {author} {\bibfnamefont {J.~C.}\ \bibnamefont
  {Slater}},\ }\bibfield  {title} {\bibinfo {title} {A simplification of the
  hartree-fock method},\ }\href {https://doi.org/10.1103/PhysRev.81.385}
  {\bibfield  {journal} {\bibinfo  {journal} {Phys. Rev.}\ }\textbf {\bibinfo
  {volume} {81}},\ \bibinfo {pages} {385} (\bibinfo {year} {1951})}\BibitemShut
  {NoStop}%
\bibitem [{\citenamefont {Harston}\ and\ \citenamefont
  {Pyper}(1992)}]{HarstonPRA1992}%
  \BibitemOpen
  \bibfield  {author} {\bibinfo {author} {\bibfnamefont {M.~R.}\ \bibnamefont
  {Harston}}\ and\ \bibinfo {author} {\bibfnamefont {N.~C.}\ \bibnamefont
  {Pyper}},\ }\bibfield  {title} {\bibinfo {title} {Exchange effects in
  \ensuremath{\beta} decays of many-electron atoms},\ }\href
  {https://doi.org/10.1103/PhysRevA.45.6282} {\bibfield  {journal} {\bibinfo
  {journal} {Phys. Rev. A}\ }\textbf {\bibinfo {volume} {45}},\ \bibinfo
  {pages} {6282} (\bibinfo {year} {1992})}\BibitemShut {NoStop}%
\bibitem [{\citenamefont {Pyper}\ and\ \citenamefont
  {Harston}(1998)}]{PyperPRSLA1998}%
  \BibitemOpen
  \bibfield  {author} {\bibinfo {author} {\bibfnamefont {N.~C.}\ \bibnamefont
  {Pyper}}\ and\ \bibinfo {author} {\bibfnamefont {M.~R.}\ \bibnamefont
  {Harston}},\ }\bibfield  {title} {\bibinfo {title} {Atomic effects on
  \ensuremath{\beta}-decay},\ }\href {https://doi.org/10.1098/rspa.1988.0128}
  {\bibfield  {journal} {\bibinfo  {journal} {Proc. R. Soc. Lond. A}\ }\textbf
  {\bibinfo {volume} {420}},\ \bibinfo {pages} {277} (\bibinfo {year}
  {1998})}\BibitemShut {NoStop}%
\bibitem [{\citenamefont {Ni\ifmmode~\mbox{\c{t}}\else \c{t}\fi{}escu}\ \emph
  {et~al.}(2023)\citenamefont {Ni\ifmmode~\mbox{\c{t}}\else \c{t}\fi{}escu},
  \citenamefont {Stoica},\ and\ \citenamefont {\ifmmode~\check{S}\else
  \v{S}\fi{}imkovic}}]{NitescuPRC2023}%
  \BibitemOpen
  \bibfield  {author} {\bibinfo {author} {\bibfnamefont {O.}~\bibnamefont
  {Ni\ifmmode~\mbox{\c{t}}\else \c{t}\fi{}escu}}, \bibinfo {author}
  {\bibfnamefont {S.}~\bibnamefont {Stoica}},\ and\ \bibinfo {author}
  {\bibfnamefont {F.}~\bibnamefont {\ifmmode~\check{S}\else
  \v{S}\fi{}imkovic}},\ }\bibfield  {title} {\bibinfo {title} {Exchange
  correction for allowed $\ensuremath{\beta}$ decay},\ }\href
  {https://doi.org/10.1103/PhysRevC.107.025501} {\bibfield  {journal} {\bibinfo
   {journal} {Phys. Rev. C}\ }\textbf {\bibinfo {volume} {107}},\ \bibinfo
  {pages} {025501} (\bibinfo {year} {2023})}\BibitemShut {NoStop}%
\bibitem [{\citenamefont {Rose}(1995)}]{RoseBook1995}%
  \BibitemOpen
  \bibfield  {author} {\bibinfo {author} {\bibfnamefont {M.~E.}\ \bibnamefont
  {Rose}},\ }\href@noop {} {\emph {\bibinfo {title} {Elementary {T}heory of
  {A}ngular {M}omentum}}}\ (\bibinfo  {publisher} {Dover},\ \bibinfo {year}
  {1995})\BibitemShut {NoStop}%
\bibitem [{\citenamefont {Varshalovich}\ \emph {et~al.}(1995)\citenamefont
  {Varshalovich}, \citenamefont {Moskalev},\ and\ \citenamefont
  {Khersonskii}}]{VarshalovichBook1988}%
  \BibitemOpen
  \bibfield  {author} {\bibinfo {author} {\bibfnamefont {D.~A.}\ \bibnamefont
  {Varshalovich}}, \bibinfo {author} {\bibfnamefont {A.~N.}\ \bibnamefont
  {Moskalev}},\ and\ \bibinfo {author} {\bibfnamefont {V.~K.}\ \bibnamefont
  {Khersonskii}},\ }\href@noop {} {\emph {\bibinfo {title} {Quantum {T}heory of
  {A}ngular {M}omentum}}}\ (\bibinfo  {publisher} {World Scientific},\ \bibinfo
  {year} {1995})\BibitemShut {NoStop}%
\bibitem [{\citenamefont {Latter}(1955)}]{LatterPR1955}%
  \BibitemOpen
  \bibfield  {author} {\bibinfo {author} {\bibfnamefont {R.}~\bibnamefont
  {Latter}},\ }\bibfield  {title} {\bibinfo {title} {Atomic energy levels for
  the thomas-fermi and thomas-fermi-dirac potential},\ }\href
  {https://doi.org/10.1103/PhysRev.99.510} {\bibfield  {journal} {\bibinfo
  {journal} {Phys. Rev.}\ }\textbf {\bibinfo {volume} {99}},\ \bibinfo {pages}
  {510} (\bibinfo {year} {1955})}\BibitemShut {NoStop}%
\bibitem [{\citenamefont {Liberman}\ \emph {et~al.}(1971)\citenamefont
  {Liberman}, \citenamefont {Cromer},\ and\ \citenamefont
  {Waber}}]{LibermanCPC1971}%
  \BibitemOpen
  \bibfield  {author} {\bibinfo {author} {\bibfnamefont {D.}~\bibnamefont
  {Liberman}}, \bibinfo {author} {\bibfnamefont {D.}~\bibnamefont {Cromer}},\
  and\ \bibinfo {author} {\bibfnamefont {J.}~\bibnamefont {Waber}},\ }\bibfield
   {title} {\bibinfo {title} {Relativistic self-consistent field program for
  atoms and ions},\ }\href
  {https://doi.org/https://doi.org/10.1016/0010-4655(71)90020-8} {\bibfield
  {journal} {\bibinfo  {journal} {Computer Physics Communications}\ }\textbf
  {\bibinfo {volume} {2}},\ \bibinfo {pages} {107} (\bibinfo {year}
  {1971})}\BibitemShut {NoStop}%
\bibitem [{\citenamefont {Liberman}\ \emph {et~al.}(1965)\citenamefont
  {Liberman}, \citenamefont {Waber},\ and\ \citenamefont
  {Cromer}}]{LibermanPR1965}%
  \BibitemOpen
  \bibfield  {author} {\bibinfo {author} {\bibfnamefont {D.}~\bibnamefont
  {Liberman}}, \bibinfo {author} {\bibfnamefont {J.~T.}\ \bibnamefont
  {Waber}},\ and\ \bibinfo {author} {\bibfnamefont {D.~T.}\ \bibnamefont
  {Cromer}},\ }\bibfield  {title} {\bibinfo {title} {Self-consistent-field
  dirac-slater wave functions for atoms and ions. i. comparison with previous
  calculations},\ }\href {https://doi.org/10.1103/PhysRev.137.A27} {\bibfield
  {journal} {\bibinfo  {journal} {Phys. Rev.}\ }\textbf {\bibinfo {volume}
  {137}},\ \bibinfo {pages} {A27} (\bibinfo {year} {1965})}\BibitemShut
  {NoStop}%
\bibitem [{\citenamefont {Moliere}(1947)}]{MoliereZNA1947}%
  \BibitemOpen
  \bibfield  {author} {\bibinfo {author} {\bibfnamefont {G.}~\bibnamefont
  {Moliere}},\ }\bibfield  {title} {\bibinfo {title} {Theorie der streuung
  schneller geladener teilchen i. einzelstreuung am abgeschirmten
  coulomb-feld},\ }\href {https://doi.org/doi:10.1515/zna-1947-0302} {\bibfield
   {journal} {\bibinfo  {journal} {Zeitschrift für Naturforschung A}\ }\textbf
  {\bibinfo {volume} {2}},\ \bibinfo {pages} {133} (\bibinfo {year}
  {1947})}\BibitemShut {NoStop}%
\bibitem [{\citenamefont {{Froese Fischer}}\ \emph {et~al.}(2019)\citenamefont
  {{Froese Fischer}}, \citenamefont {Gaigalas}, \citenamefont {Jönsson},\ and\
  \citenamefont {Bieroń}}]{FroeseFischerCPC2018}%
  \BibitemOpen
  \bibfield  {author} {\bibinfo {author} {\bibfnamefont {C.}~\bibnamefont
  {{Froese Fischer}}}, \bibinfo {author} {\bibfnamefont {G.}~\bibnamefont
  {Gaigalas}}, \bibinfo {author} {\bibfnamefont {P.}~\bibnamefont {Jönsson}},\
  and\ \bibinfo {author} {\bibfnamefont {J.}~\bibnamefont {Bieroń}},\
  }\bibfield  {title} {\bibinfo {title} {Grasp2018—a fortran 95 version of
  the general relativistic atomic structure package},\ }\href
  {https://doi.org/https://doi.org/10.1016/j.cpc.2018.10.032} {\bibfield
  {journal} {\bibinfo  {journal} {Computer Physics Communications}\ }\textbf
  {\bibinfo {volume} {237}},\ \bibinfo {pages} {184} (\bibinfo {year}
  {2019})}\BibitemShut {NoStop}%
\bibitem [{\citenamefont {Carlson}(1975)}]{CarlsonBook1975}%
  \BibitemOpen
  \bibfield  {author} {\bibinfo {author} {\bibfnamefont {T.~A.}\ \bibnamefont
  {Carlson}},\ }\href@noop {} {\emph {\bibinfo {title} {Photoelectron and
  {A}uger {S}pectroscopy}}}\ (\bibinfo  {publisher} {Plenum Press},\ \bibinfo
  {year} {1975})\BibitemShut {NoStop}%
\bibitem [{\citenamefont {Haselschwardt}\ \emph {et~al.}(2020)\citenamefont
  {Haselschwardt}, \citenamefont {Kostensalo}, \citenamefont {Mougeot},\ and\
  \citenamefont {Suhonen}}]{Haselschwardt-PRC2020}%
  \BibitemOpen
  \bibfield  {author} {\bibinfo {author} {\bibfnamefont {S.~J.}\ \bibnamefont
  {Haselschwardt}}, \bibinfo {author} {\bibfnamefont {J.}~\bibnamefont
  {Kostensalo}}, \bibinfo {author} {\bibfnamefont {X.}~\bibnamefont
  {Mougeot}},\ and\ \bibinfo {author} {\bibfnamefont {J.}~\bibnamefont
  {Suhonen}},\ }\bibfield  {title} {\bibinfo {title} {Improved calculations of
  $\ensuremath{\beta}$ decay backgrounds to new physics in liquid xenon
  detectors},\ }\href {https://doi.org/10.1103/PhysRevC.102.065501} {\bibfield
  {journal} {\bibinfo  {journal} {Phys. Rev. C}\ }\textbf {\bibinfo {volume}
  {102}},\ \bibinfo {pages} {065501} (\bibinfo {year} {2020})}\BibitemShut
  {NoStop}%
\bibitem [{\citenamefont {Basunia}(2009)}]{BasuniaNDS2009}%
  \BibitemOpen
  \bibfield  {author} {\bibinfo {author} {\bibfnamefont {M.}~\bibnamefont
  {Basunia}},\ }\bibfield  {title} {\bibinfo {title} {Nuclear data sheets for
  {A} = 187},\ }\href
  {https://doi.org/https://doi.org/10.1016/j.nds.2009.04.001} {\bibfield
  {journal} {\bibinfo  {journal} {Nuclear Data Sheets}\ }\textbf {\bibinfo
  {volume} {110}},\ \bibinfo {pages} {999} (\bibinfo {year}
  {2009})}\BibitemShut {NoStop}%
\bibitem [{\citenamefont {Mougeot}(2015)}]{MougeotPRC2015}%
  \BibitemOpen
  \bibfield  {author} {\bibinfo {author} {\bibfnamefont {X.}~\bibnamefont
  {Mougeot}},\ }\bibfield  {title} {\bibinfo {title} {Reliability of usual
  assumptions in the calculation of $\ensuremath{\beta}$ and $\ensuremath{\nu}$
  spectra},\ }\href {https://doi.org/10.1103/PhysRevC.91.055504} {\bibfield
  {journal} {\bibinfo  {journal} {Phys. Rev. C}\ }\textbf {\bibinfo {volume}
  {91}},\ \bibinfo {pages} {055504} (\bibinfo {year} {2015})}\BibitemShut
  {NoStop}%
\bibitem [{\citenamefont {Caprio}(2005)}]{SciDraw}%
  \BibitemOpen
  \bibfield  {author} {\bibinfo {author} {\bibfnamefont {M.}~\bibnamefont
  {Caprio}},\ }\bibfield  {title} {\bibinfo {title} {Levelscheme: A level
  scheme drawing and scientific figure preparation system for mathematica},\
  }\href {https://doi.org/https://doi.org/10.1016/j.cpc.2005.04.010} {\bibfield
   {journal} {\bibinfo  {journal} {Computer Physics Communications}\ }\textbf
  {\bibinfo {volume} {171}},\ \bibinfo {pages} {107 } (\bibinfo {year}
  {2005})}\BibitemShut {NoStop}%
\end{thebibliography}%
	
\end{document}